\documentclass[12pt,preprint]{aastex}    
\usepackage{graphicx,amssymb,enumerate}

\newcommand{\kms}{km s$^{-1}$}
\newcommand{\aforty}{$\alpha$.40}
\newcommand{\Jykms}{Jy$\:$km$\,$s$^{-1}$}

\newcommand{\rband}{$r$-band}
\newcommand{\LCDM}{$\Lambda$CDM}
\newcommand{\Xis}{$\Xi(\sigma) \, / \, \sigma$}
\newcommand{\xisp}{$\xi(\sigma,\pi)$}

\shorttitle{Clustering properties of ALFALFA galaxies}
\shortauthors{Papastergis et al.}

\begin{document}

\title{The clustering of ALFALFA galaxies: dependence on HI mass, relationship to optical samples \& clues on host halo properties.}

\author {Emmanouil Papastergis\altaffilmark{1}, Riccardo Giovanelli\altaffilmark{1}, Martha P. Haynes\altaffilmark{1}, Aldo Rodr\'iguez--Puebla\altaffilmark{2} \& Michael G. Jones\altaffilmark{1}}



\altaffiltext{1}{Center for Radiophysics and Space Research, Space Sciences Building,
Cornell University, Ithaca, NY 14853, USA. {\textit{e-mail:}} papastergis@astro.cornell.edu, riccardo@astro.cornell.edu, haynes@astro.cornell.edu, jonesmg@astro.cornell.edu}

\altaffiltext{2}{Instituto de Astronom\'ia, Universidad Nacional Aut\'onoma de M\'exico, A. P. 70-264, 04510, M\'exico, D.F., M\'exico. {\textit{e-mail:}} apuebla@astro.unam.mx}


\begin{abstract}

We use a sample of $\approx$$6\,000$ galaxies detected by the Arecibo Legacy Fast ALFA (ALFALFA) 21cm survey, to measure the clustering properties of HI-selected galaxies. We find no convincing evidence for a dependence of clustering on the galactic atomic hydrogen (HI) mass, over the range $M_{HI} \approx 10^{8.5} - 10^{10.5} \; M_\odot$. We show that previously reported results of weaker clustering for low-HI mass galaxies are probably due to finite-volume effects. In addition, we compare the clustering of ALFALFA galaxies with optically selected samples drawn from the Sloan Digital Sky Survey (SDSS). We find that HI-selected galaxies cluster more weakly than even relatively optically faint galaxies, when no color selection is applied. Conversely, when SDSS galaxies are split based on their color, we find that the correlation function of blue optical galaxies is practically indistinguishable from that of HI-selected galaxies. At the same time, SDSS galaxies with red colors are found to cluster significantly more than HI-selected galaxies, a fact that is evident in both the projected as well as the full two-dimensional correlation function. A cross-correlation analysis further reveals that gas-rich galaxies ``avoid'' being located within $\approx$3 Mpc of optical galaxies with red colors. Next, we consider the clustering properties of halo samples selected from the Bolshoi \LCDM \ simulation. A comparison with the clustering of ALFALFA galaxies suggests that galactic HI mass is not tightly related to host halo mass, and that a sizable fraction of subhalos do not host HI galaxies. Lastly, we find that we can recover fairly well the correlation function of HI galaxies by just excluding halos with low spin parameter. This finding lends support to the hypothesis that halo spin plays a key role in determining the gas content of galaxies.


\end{abstract}

\maketitle

\section{Introduction}

In the currently accepted hierarchical theory of structure formation, the clustering of galaxies is jointly determined by the large-scale structure of dark matter in the universe, as well as the way in which baryons trace dark matter through the formation of galaxies \citep[see e.g.][]{CooraySheth2002}. As a result, the quantitative study of galaxy clustering through the correlation function, $\xi(r)$, has been instrumental both for constraining cosmological models \citep[e.g.][]{Jenkins1998} as well as for furthering our understanding of galaxy formation and evolution \citep[e.g.][]{Zheng2007}.

Given a cosmological model, the clustering of galaxies can be used to constrain the galaxy-halo connection, thus testing and informing models of galaxy formation. A large quantity of work has been devoted to studying the correlation function of galaxies as a function of their optical properties, such as luminosity, color, morphological and spectral type (e.g. \citealp{Zehavi2005,Zehavi2011, Li2012}). These studies have established with increasing precision a number of fundamental clustering phenomena, such as the trend for stronger clustering with increasing luminosity and the fact that galaxies with blue colors, late-type morphologies and elevated star formation activity cluster significantly less than red, early-type, quiescent galaxies. Moreover, several studies have used the halo occupation distribution (HOD) formalism to make quantitative predictions for the properties of halos hosting a certain class of galaxies \citep[e.g.][]{Zehavi2005}. These analyses have suggested that more luminous galaxies inhabit more massive halos on average, and that red galaxies have a higher chance of being hosted by a subhalo compared with blue galaxies. These results are in agreement with theoretical expectations, and are supported by a number of other observational methods (e.g. abundance matching: \citealp{Guo2010, Behroozi2010, Moster2010, Leauthaud2012} and galaxy-galaxy weak lensing: \citealp{Mandelbaum2006, Dutton2010, Reyes2012}). These clustering based galaxy occupation models then feed back into cosmological studies, since they provide the necessary link between the measured distribution of galaxies and the distribution of matter that is determined by the cosmological parameters \citep[e.g.][]{Reddick2012}.

Until recently, similarly detailed studies of the clustering characteristics of galaxies selected by their atomic hydrogen content (HI-selected) were not feasible, due to the lack of large-volume blind 21cm surveys. In recent years however, the HI Parkes All Sky Survey (HIPASS; \citealp{Meyer2004}) and the Arecibo Legacy Fast ALFA\footnotemark{} (ALFALFA; \citealp{Giovanelli2005}) survey have provided adequate samples for this purpose. \citet{Basilakos2007} and \citet{Meyer2007} have both analyzed the HIPASS dataset, establishing the fact that HI-selected galaxies are among the most weakly clustered galaxy populations known. In addition, both of these works investigated the dependence of clustering strength on galaxy HI mass ($M_{HI}$), arriving at different conclusions. More recently, \citet{Martin2012} used $\approx$$10\,000$ galaxies from the 40\% ALFALFA catalog (``\aforty'' catalog; \citealp{Haynes2011}) to measure the correlation function of gas-rich galaxies. Among their main findings were that HI-selected galaxies show a markedly anisotropic clustering pattern (their Fig. 3, see also Fig.\ref{fig:corr2d_hi} in this article), and that they are anti-biased with respect to dark matter on scales $\lesssim 5$ Mpc (their Fig. 10). 


\footnotetext{The Arecibo L-band Feed Array (ALFA) is a 7-feed receiver operating in the L-band ($\approx 1420$ MHz), installed at the Arecibo Observatory.}

In this work, we take advantage of the large HI dataset provided by ALFALFA to make a detailed investigation of the clustering properties of gas-rich galaxies. We furthermore draw samples from the spectroscopic database of the 7th data release of the SDSS \citep{Abazajian2009} spanning the same volume as the ALFALFA sample, to make comparisons with the clustering properties of optically selected galaxies. The fact that the ALFALFA and SDSS samples are drawn from the same volume allows for a further cross-correlation analysis, measuring the spatial relationship between HI and optical galaxies. Lastly, we select halos from the Bolshoi \LCDM \ simulation \citep{Klypin2011}, to investigate what halo properties are associated with weak clustering, giving us evidence on the characteristics of halos hosting gas-rich galaxies.

The paper is organized as follows: in section \ref{sec:data_methods}, we present the ALFALFA and SDSS samples used to measure the clustering of HI and optical galaxies, and we describe the methodology for measuring the correlation function. In section \ref{sec:results} we present our results concerning the clustering properties of a number of HI-selected and optically selected samples, and discuss the implications. In section \ref{sec:halos} we present our halo samples selected from the Bolshoi simulation, and study their clustering as a function of their properties (mass, spin, etc.). We conclude in section \ref{sec:conclusion}, by summarizing the main findings of this work. We forewarn the reader that --unlike most correlation function articles-- all distances in this work assume a Hubble constant of $H_0 = 70 \, h_{70} \; \mathrm{km}\, \mathrm{s}^{-1}  \mathrm{Mpc}^{-1}$. In order to facilitate comparisons with the literature however, the upper $x$-axis of Figures is expressed in terms of $h \equiv H_0 / 100 \; \mathrm{km}\, \mathrm{s}^{-1}  \mathrm{Mpc}^{-1}$ when appropriate.


\section{Data \& Methods}
\label{sec:data_methods}

\subsection{ALFALFA sample}
\label{sec:alfa_sample}

The ALFALFA survey is a wide-area, blind 21cm emission-line survey performed with the 305m radio telescope at the Arecibo Observatory \citep{Giovanelli2005}. The survey has recently completed data acquisition, and a source catalog covering $\approx$40\% of the final survey area has been publicly released (``\aforty'' catalog; \citealp{Haynes2011}). ALFALFA has greater sensitivity, finer spectral resolution and better centroiding accuracy than previous blind HI surveys of comparable sky coverage (e.g. HIPASS; \citealp{Barnes2001}), and \aforty \ already represents the largest HI-selected galaxy sample to date.

\begin{figure}[htbp]
\centering
\includegraphics[scale=0.6]{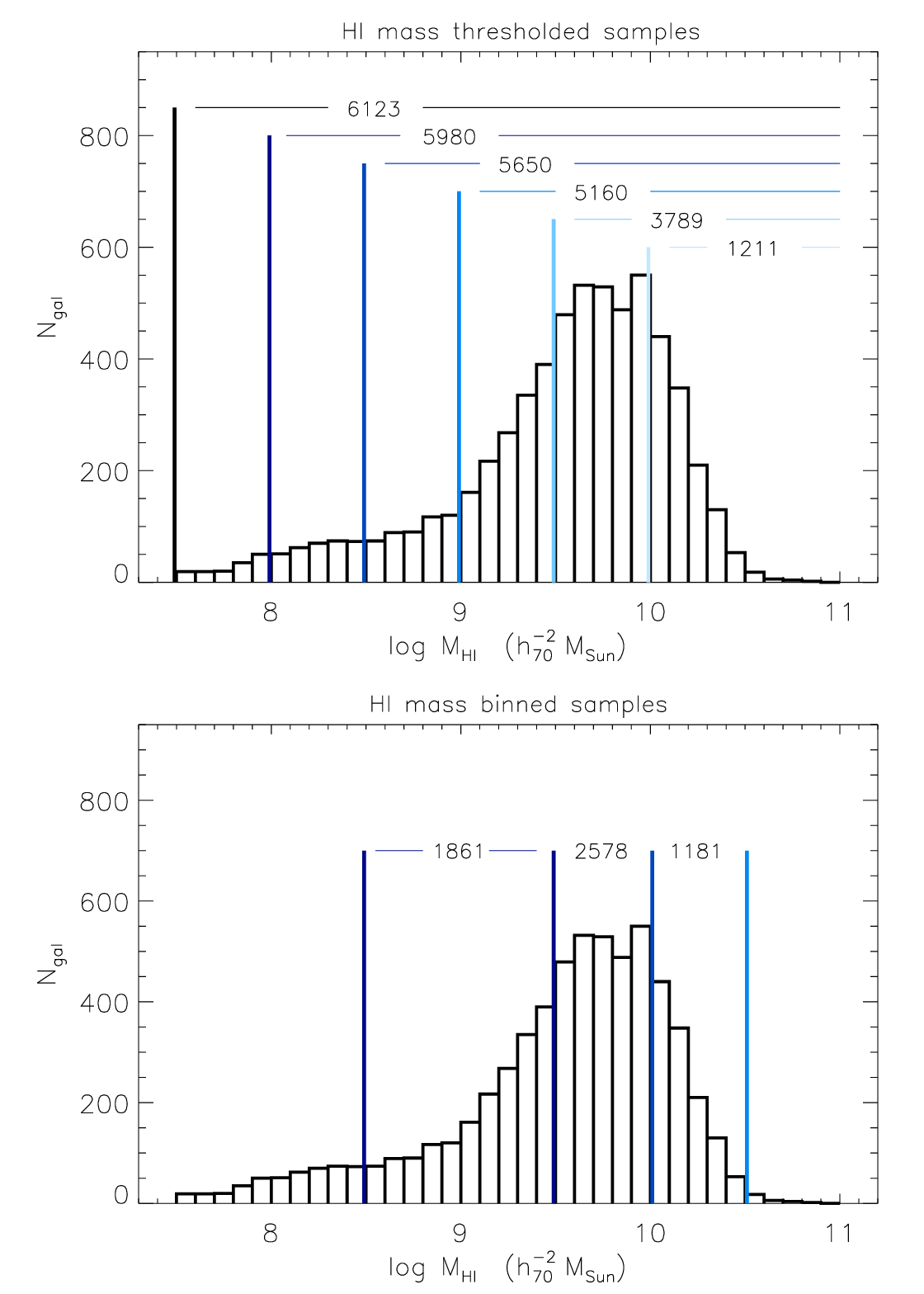}
\caption{ \footnotesize{The histogram represents the ALFALFA parent sample counts in bins of HI mass. The upper panel is a graphical representation of the HI mass thresholded samples, where each colored vertical line denotes the HI mass threshold of the sample and the total number of galaxies in each sample is quoted. The lower panel is the corresponding plot for the HI mass binned samples. Vertical colored lines denote the HI mass bins limits, while the number of galaxies in each binned sample is also quoted.} }
\label{fig:himass_hist}
\end{figure}


In this article, we use a parent sample of 6$\,$123 HI-selected galaxies detected by the ALFALFA survey. In particular, we select galaxies over a contiguous rectangular sky region of $\approx 1\,700$ deg$^2$ ($135^\circ < \mathrm{RA} < 230^\circ$ and $0^\circ < \mathrm{Dec} < 18^\circ$) and in the redshift range $z \approx 0.0023 - 0.05$ ($\mathrm{v}_{\mathrm{CMB}} = 700 - 15000$ \kms). The parent sample has significant overlap with the publicly available \aforty \ sample, but has been supplemented by newly processed ALFALFA regions covering the declination ranges $0^\circ < \mathrm{Dec} < 4^\circ$ \& $16^\circ < \mathrm{Dec} < 18^\circ$. The sample is restricted to ``Code 1'' ALFALFA detections, i.e. it is comprised only by confidently detected extragalactic sources ($S/N_{HI} > 6.5$). In addition, parent sample sources have a combination of observed 21cm flux ($S_{HI}$) and 21cm lineprofile width ($W_{50}$) that places them in the region of the $\{S_{HI},W_{50}\}$--plane where the completeness of the ALFALFA survey is at least 50\% (see Sec.  6 and Fig.~12 in \citealt{Haynes2011}). Lastly, the sample is limited to linewidths $W_{50} > 18$ \kms \ and HI masses\footnotemark{} $M_{HI} > 10^{7.5} M_\odot$.

\footnotetext{Atomic hydrogen (HI) masses for ALFALFA galaxies are calculated from their 21cm line flux though the relation $M_{HI} = 2.356\,10^5 \; S_{HI} \; d^2$. In this formula $M_{HI}$ is measured in $M_\odot$ and the flux $S_{HI}$ in \Jykms. The distance $d$ is measured in Mpc, and calculated from the galaxy's recessional velocity in the CMB frame as $d=\mathrm{v}_{CMB}/H_0$.}

From the parent sample described above we select a number of subsamples, such that their HI mass is above a specified limit (HI mass thresholds) or within a specific range (HI mass bins). Figure~\ref{fig:himass_hist} shows a histogram of $M_{HI}$ for the parent sample, with a graphical representation of the HI mass-thresholded and HI mass-binned subsamples used in this work.

\subsection{SDSS sample}
\label{sec:sdss_sample}

We select an optical sample of galaxies from the spectroscopic database of the 7th data release of the Sloan Digital Sky Survey (SDSS DR7; \citealp{Abazajian2009}). This optically selected parent sample is restricted to the same volume as the HI-selected sample used in this work:  $135^\circ < \mathrm{RA} < 230^\circ$, $0^\circ < \mathrm{Dec} < 18^\circ$ and $z \approx 0.0023 - 0.05$ ($\mathrm{v}_{\mathrm{CMB}} = 700 - 15000$ \kms). We only select SDSS galaxies that are spectroscopically classified as galaxies (\texttt{SpecClass = 2}) and that have an apparent magnitude in the \rband \ brighter than 17.6, after correction for Milky Way extinction ($m_r < 17.6$). In addition, we impose a color cut on our spectroscopic sources, $(i-z)_{model} > -0.25$, which excludes a small number of objects; the vast majority of them are cases where star-forming knots and structures in nearby extended spirals or dwarf irregular galaxies are erroneously classified as separate galaxies by the SDSS pipeline. Lastly, our SDSS parent sample is limited to MW extinction-corrected absolute magnitudes in the \rband \ brighter than -17 ($M_r < -17$), and is comprised by a total of $18\,516$ galaxies.

\begin{figure}[htbp]
\centering
\includegraphics[scale=0.6]{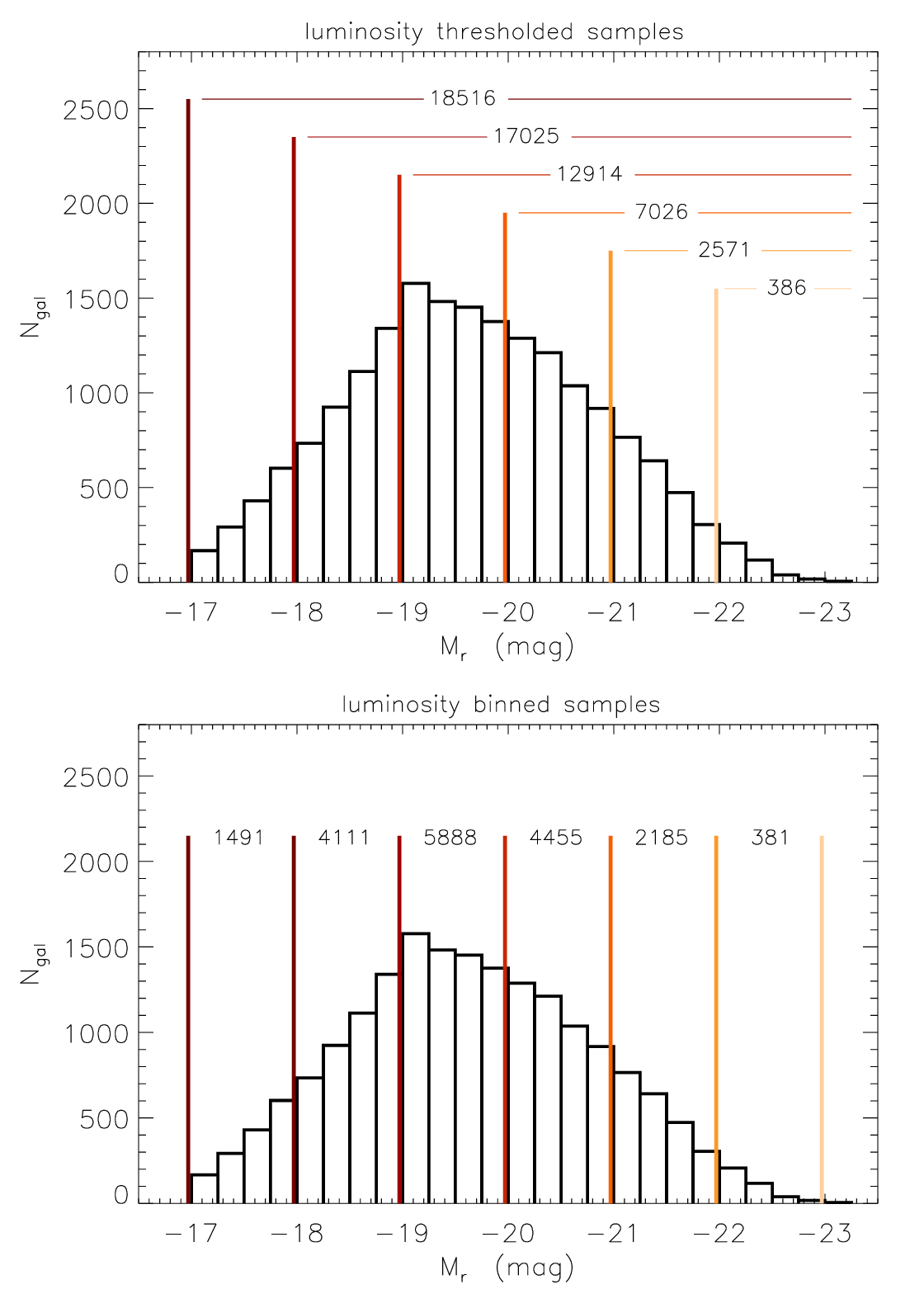}
\caption{ \footnotesize{Similar to Fig.~\ref{fig:himass_hist}, but for the SDSS parent sample. The histogram is the sample count in bins of $r$-band absolute magnitude, while the upper and lower panels represent graphically the luminosity thresholded and luminosity binned SDSS samples.} }
\label{fig:mag_hist}
\end{figure}

\begin{figure}[htbp]
\centering
\includegraphics[scale=0.6]{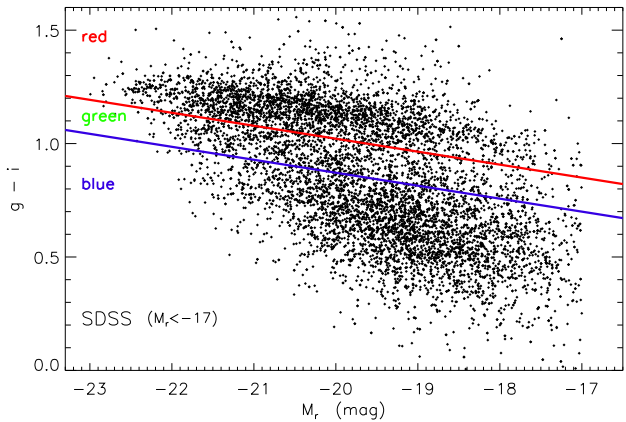}
\caption{ \footnotesize{Color-magnitude diagram (CMD) of the SDSS parent sample (only a representative subsample is plotted, for visual clarity). The horizontal axis is galactic  $r$-band absolute magnitude, while the vertical axis is the galaxy $g-i$ color. Both quantities are corrected for MW extinction. The solid lines denote the cuts used to select the ``red'', ``green'' and ``blue'' SDSS galaxy samples. The upper red boundary line is given by $g-i = 0.0571(M_r+24)+1.25 $, while the lower blue boundary line is parallel to the former with a 0.15 mag color offset from it.} }
\label{fig:cmd}
\end{figure}

From this optical parent sample we create subsamples, selected based on specifying their faintest \rband \ absolute magnitude (magnitude thresholds) or their range of \rband \ absolute magnitudes (magnitude bins). Figure~\ref{fig:mag_hist} shows a histogram of $M_r$ for the parent sample, with a graphical representation of the magnitude-thresholded and magnitude-binned subsamples used in this work. Furthermore, we define three color-based subsamples according to the position of galaxies in a color-magnitude diagram (CMD), as shown in Figure~\ref{fig:cmd}. The ``red'' subsample is composed by red sequence galaxies, the ``blue'' subsample by blue cloud galaxies and the green subsample by galaxies with intermediate locations on the CMD, sometimes referred to as ``green valley'' galaxies.

\subsection{Sample selection functions \& random catalogs} 

\begin{figure}[htbp]
\centering
\includegraphics[scale=0.55]{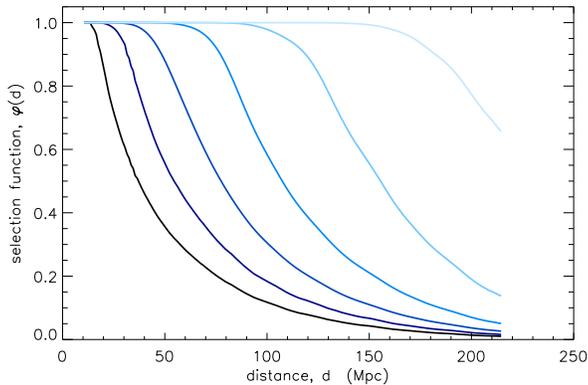}
\caption{ \footnotesize{Selection functions for the HI mass thresholded samples of Fig.~\ref{fig:himass_hist} (solid lines from bottom to top correspond to samples with $M_{HI} > 10^{7.5}, 10^8, 10^{8.5}, 10^9,  10^{9.5},  10^{10} \; M_\odot$). These are the selection functions that are used for constructing the random catalog corresponding to each of the HI mass thresholded samples.} }
\label{fig:sel_fun}
\end{figure}

Measuring the clustering of galaxies with certain properties involves comparing the spatial distribution of an observed galactic sample with the spatial distribution of a catalog of random points, which reflect the galactic sample's selection function \citep[see e.g.][]{Peebles1980}. The selection function, $\varphi(d)$, describes the fraction of a hypothetical volume-limited sample of galaxies with the desired properties that is included in an observational sample at distance $d$ from the observer. For example, Figure~\ref{fig:sel_fun} shows $\varphi(d)$ for the HI mass-thresholded samples used in this work; samples restricted to more massive galaxies are complete (i.e. $\varphi=1$) out to larger distances.

Deriving the selection function for a sample is not straightforward, and necessitates two inputs: $i)$ the cuts used to define an observational sample and $ii)$ the intrinsic distribution of the galaxy properties that determine the inclusion of a galaxy in the observational sample. The SDSS sample described in \S\ref{sec:sdss_sample} is mostly flux-limited, because included galaxies satisfy an apparent \rband \ magnitude cut, $m_r < 17.6$. We thus need to calculate the intrinsic distribution of \rband \ luminosity for SDSS galaxies, most commonly referred to as the galactic luminosity function (LF). Then $\varphi(d)$ can be calculated in terms of the galaxy luminosity function, $n(M_r)$, as

\begin{equation}
\label{eqn:sf_sdss}
\varphi(d) = \frac{ \int_{M_{r,lim}(d)}^{M_{r,min}} \; n(M_r) \, dM_r }{ \int_{M_{r,max}}^{M_{r,min}} \; n(M_r) \, dM_r } \; \; .
\end{equation}

\begin{figure}[htbp]
\centering
\includegraphics[scale=0.65]{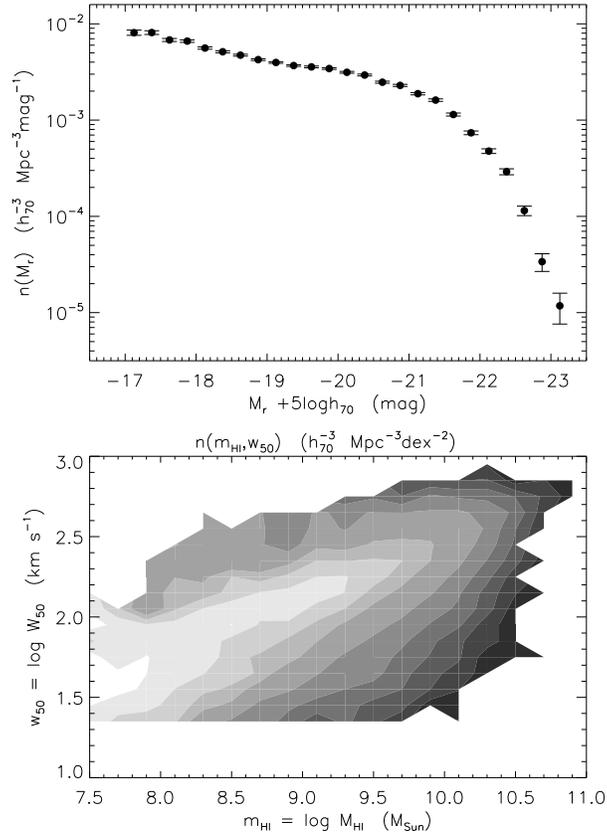}
\caption{ \footnotesize{\textit{upper panel:} The SDSS $r$-band luminosity function, used to calculate the selection function for the SDSS subsamples used in this work. The error bars denote only the Poisson error due to the number of sample galaxies in each $M_r$ bin. \textit{lower panel:} The ALFALFA mass-width function, used to calculate the selection function for the ALFALFA subsamples used in this work. The contours are set at $n(m_{HI},w_{50}) = 10^{-6}, 10^{-5.5}, \ldots,10^{-2.5},10^{-1.75},\ldots,10^{-1.25} \;\; \mathrm{Mpc}^{-3}\mathrm{dex}^{-2}$, from darker to lighter tones.} }
\label{fig:lf+mwf}
\end{figure}

\noindent
$M_{r,max}$ and $M_{r,min}$ are the faint and bright absolute magnitude limits defining a specific subsample, while $M_{r,lim}(d)$ is the faintest absolute magnitude that a galaxy at distance $d$ can have and still have an apparent magnitude brighter than $m_r = 17.6$. The luminosity function, $n(M_r)$, is the volume-limited number density of galaxies within a bin of magnitude centered on $M_r$, and has units of $\mathrm{Mpc}^{-3} \, \mathrm{mag}^{-1}$. In informal terms, the denominator in Eqn.~\ref{eqn:sf_sdss} represents the volume-limited number density of a specific subsample, while the numerator represents the number density of galaxies in the subsample that are detectable at distance $d$.  


On the other hand the ALFALFA sample described in \S\ref{sec:alfa_sample} is not a purely flux-limited sample, but it is  mostly defined through a flux-width--dependent cut (Eqns. 4 \& 5 in Sec. 6 of \citealp{Haynes2011}). We therefore need to know the intrinsic two-dimensional distribution of HI mass and linewidth, $n(m_{HI},w_{50})$, of ALFALFA galaxies. The mass-width-function is customarily expressed in logarithmic intervals of mass and width, so $m_{HI} = \log(M_{HI}/M_\odot)$ and $w_{50} = \log(W_{50}/ \mathrm{km}\,\mathrm{s}^{-1})$. The selection function for any ALFALFA subsample is then given by the expression:

\begin{equation}
\label{eqn:sf_alfa}
\varphi(d) = \frac{\int_{w_{50,min}}^{w_{50,max}} \int_{m_{HI,lim}(d,w_{50})}^{m_{HI,max}} \; n(m_{HI},w_{50}) \; dm_{HI} \, dw_{50} }{ \int_{w_{50,min}}^{w_{50,max}} \int_{m_{HI,min}}^{m_{HI,max}} \; n(m_{HI},w_{50}) \; dm_{HI} \, dw_{50} } \; \; .
\end{equation}

\noindent 
Again, $m_{HI,min}$, $m_{HI,max}$, $w_{50,min}$ and $w_{50,max}$ are the HI mass and linewidth limits defining a specific ALFALFA subsample, while $m_{HI,lim}(d,w_{50})$ is the minimum HI mass detectable at distance $d$ for a source of linewidth $w_{50}$, as dictated by the ALFALFA 50\% completeness limit. $n(m_{HI},w_{50})$ is again the number density of galaxies within a logarithmic bin of HI mass centered on $m_{HI}$ and a logarithmic bin of width centered on $w_{50}$, and has units of $\mathrm{Mpc}^{-3} \, \mathrm{dex}^{-2}$.

\begin{figure}[htbp]
\centering
\includegraphics[scale=0.5]{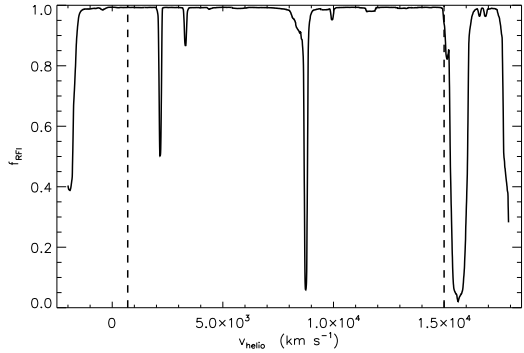}
\caption{ \footnotesize{The black solid line represents the fraction of the nominal surveyed volume available for ALFALFA source extraction in the presence of radio frequency interference (RFI), as a function of heliocentric velocity (roughly equivalent to antenna rest frequency). The largest dip at $v_\odot \approx 16\,000$ \kms \ is due to the San Juan airport radar, while the second largest dip at $v_\odot \approx 8\,800$ \kms \ is one of the radar's harmonics. The vertical dashed lines are the approximate redshift limits of the ALFALFA parent sample used in this article; the high redshift limit has been deliberately chosen so as to avoid the strongest RFI peak.} }
\label{fig:rfi}
\end{figure}

The \rband \ luminosity function for SDSS galaxies and the mass-width-function for ALFALFA galaxies used in Eqns.~\ref{eqn:sf_sdss} \& \ref{eqn:sf_alfa} are shown in Figure~\ref{fig:lf+mwf}. We calculate them by applying appropriate volume correction factors to the sample histograms of $M_r$ and $\{m_{HI},w_{50}\}$, respectively. We use the maximum-likelihood, non-parametric ``$1/V_{eff}$'' method \citep{Zwaan2005} to calculate the volume weights, on a galaxy-by-galaxy basis. For a more detailed description of the application of the method on the SDSS and ALFALFA datasets see \S3.1 in \citet{Papastergis2012} and referenced articles therein.


\begin{figure}[htbp]
\centering
\includegraphics[scale=0.50]{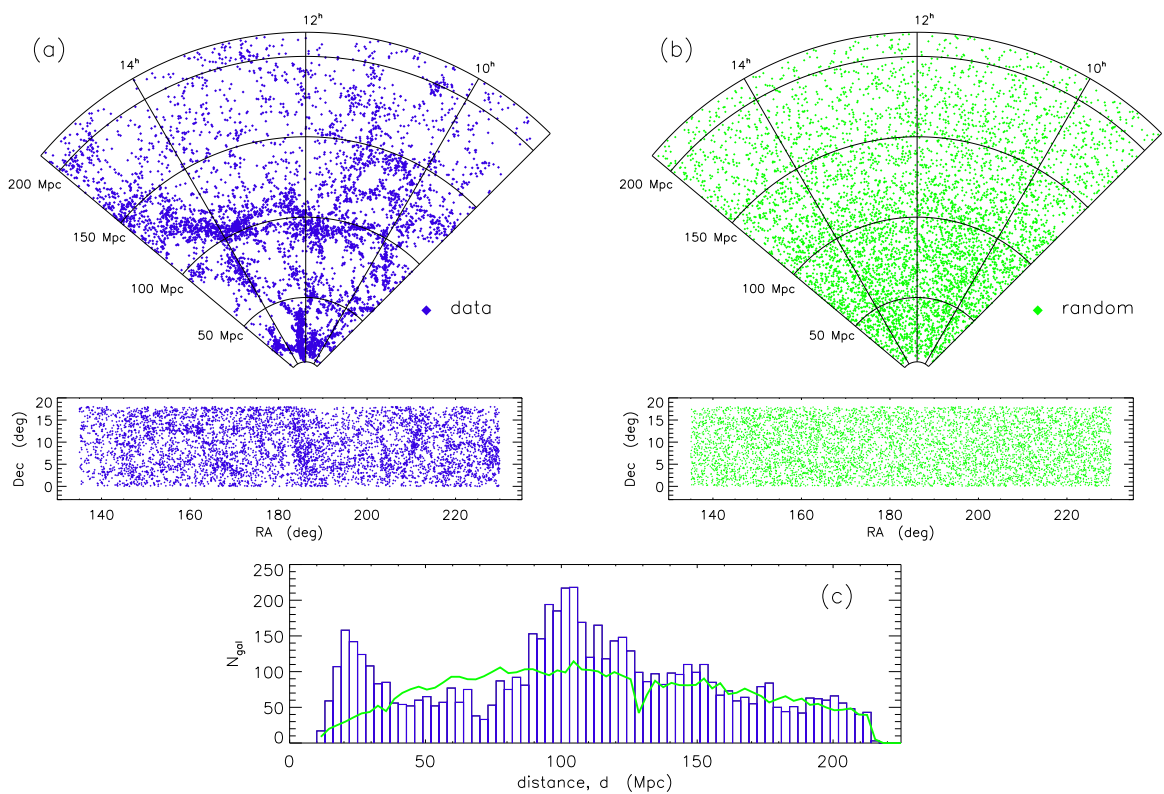}
\caption{ \footnotesize{\textit{panel (a):} Coneplot (polar plot of distance \& RA) and sky position (rectangular plot of RA \& Dec) of the ALFALFA parent sample (blue diamonds). \textit{panel (b):} Same as panel (a), but for the corresponding catalog of random points (green diamonds). \textit{panel (c):} Distance distribution of the ALFALFA parent sample (blue histogram) and its corresponding random catalog (green line). Note the markedly non-uniform distribution of ALFLAFA galaxies, which is evident in all three plots.} }
\label{fig:data+random}
\end{figure}



Once a data sample selection function is known, it is straightforward to construct a random catalog suitable for the calculation of the sample correlation function. Initially, random points are created within the subsample volume with a constant expected number density throughout, $\left<dN_{rand}/dV\right> = \mathrm{const}$. This translates into random points being uniformly distributed in RA, $\sin(\mathrm{Dec})$ and $d^3$. Subsequently, each random point is kept with probability $\varphi(d)$ (where $d$ is the random point's distance from the observer), in order to reproduce the subsample's selection function. In the case of an ALFALFA sample an additional step is necessary: accounting for the effects of radio frequency interference (RFI). RFI disrupts ALFALFA's performance in the frequency bands where it occurs, resulting in galaxies with certain heliocentric velocities having a lower chance  of being detected. Figure~\ref{fig:rfi} shows the fractional ALFALFA volume lost to RFI as a function of heliocentric velocity. In order to reproduce the effects of RFI on the spatial distribution of ALFALFA samples, points in the random catalog are kept with a probability $f_{RFI}(\mathrm{v}_\odot)$ where $\mathrm{v}_\odot$ is the heliocentric velocity of the random point\footnotemark{}. 

Figure \ref{fig:data+random} compares the distribution of data and random points for the $M_{HI} = 10^{9.5} - 10^{10} \, M_\odot$ ALFALFA sample. Panel (a) displays the coneplot (i.e. a projection of $\mathrm{RA}$ and $d$ in polar coordinates) and the sky distribution  of the data sample, while panel (b) displays the same distributions for the corresponding random catalog. Panel (c) compares the distance histograms of the two samples. The non-uniform distribution of the data set and the large-scale structure in the survey volume are readily visible in all three panels.

\footnotetext{The heliocentric velocity of a random point is calculated by first considering their velocity relative to the cosmic microwave background (CMB), $\mathrm{v}_{CMB} = H_0 \, d$, and then converting the velocity from the CMB to the heliocentric frame according to the random point's position in the sky.}

\subsection{Clustering measures}
\label{sec:clustering_measures}

The galaxy correlation function at a given length scale, $\xi(r)$, is defined as the excess probability of finding a pair of galaxies separated by distance $r$ compared to the case of a randomly distributed set of points. It then follows that a positive value of $\xi(r)$ means that the sample under consideration tends to cluster on length scales $r$, while a negative value means that it avoids clustering on this scale (a randomly distributed sample would have $\xi(r)=0$ for all $r$). In formal terms, the correlation function is defined through the relation $ \left<d^2 N_{pair}\right> = \bar{n}_{gal}^2 \; (1+\xi(r)) \; dV_1 \, dV_2$. Here $dV_1$ and $dV_2$ are two volume elements separated by distance $r$ and $\bar{n}_{gal}$ is the average galaxy number density. $\left<d^2N_{pair}\right>$ is the average number of galaxy pairs within those volume elements, which would be just $\bar{n}_{gal}^2 \, dV_1 \, dV_2$ in the absence of clustering. In practice, when a galactic sample and its corresponding random catalog are available, the correlation function is calculated in terms of the number of data-data, random-random and data-random pairs whose separation falls in the bin $r \pm \Delta r/2$. These pair counts are denoted by $P_{DD}(r), P_{RR}(r)$ \& $P_{DR}(r)$, respectively. If the data sample contains $N_D$ objects and the random sample $N_R$ objects, we can compute the normalized counts

\begin{eqnarray}
DD(r) & =& P_{DD}(r) \, / \, (N_D(N_D-1)/2)  \nonumber \\ 
RR(r) &=& P_{RR}(r) \, / \, (N_R(N_R-1)/2)  \\
DR(r) &=& P_{DR}(r) \, / \, (N_D N_R) \;\; , \nonumber 
\end{eqnarray}

\noindent
where in all three cases the denominator represents the total number of available pairs. 

The most intuitive estimator for the correlation function is then $\hat{\xi}(r) = DD(r)  /  RR(r) - 1$, which just computes the ratio of the fraction of data-data pairs and random-random pairs separated by distance $r$ and compares it with unity. However, \citet{LS1993} have shown that an alternative estimator, $\hat{\xi}_{LS}(r) = (DD(r)  -2 DR(r) + RR(r)) \, / \, RR(r)$, has better statistical performance; for volume-limited, weakly clustered samples equipped with large random catalogs ($N_R \gg N_D$) the $\hat{\xi}_{LS}$ estimator is unbiased and its variance is determined just by the counting noise associated with the number of data-data pairs. In this article we adopt throughout the LS estimator, dropping from now on the excess notation: $\hat{\xi}_{LS}(r) \rightarrow \xi(r)$.    

Despite the fact that the ``real space'' correlation function, $\xi(r)$, is the fundamental quantity related to galaxy clustering, physical separation is not generally available for extragalactic objects. The measurable quantities in a spectroscopic galaxy survey are position on the sky $(\mathrm{RA}, \mathrm{Dec})$ and recessional velocity ($\mathrm{v}_{CMB} = c \, z$). As a result we consider in this article the ``redshift space'' separation between two objects $s$, given by

\begin{equation}
s = \sqrt{(v_1^2 + v_2^2-2v_1v_2\cos\theta)}/H_0 \; \; ,
\end{equation}

\noindent
where $v_1, v_2$ are the recessional velocities of galaxies 1 and 2 respectively in \kms, $\theta$ is the angle between them on the sky, and $H_0$ is the Hubble constant (recall that in this article $H_0 = 70$ \kms \ Mpc$^{-1}$). In addition we can consider separately the components of the separation along the line of sight ($\pi$) and on the plane of the sky ($\sigma$) defined as:

\begin{eqnarray}
\pi &=& |v_1-v_2| \, / \, H_0 \;\;\;\; \mathrm{and} \\
\sigma &=& \sqrt{s^2 - \pi^2} \; .
\end{eqnarray}   

\noindent 
We can therefore calculate the redshift space correlation function $\xi(s)$ by counting the number of pairs whose separation is within $s \pm \Delta s / 2$. Similarly, we can calculate the two-dimensional correlation function $\xi(\sigma,\pi)$ by counting pairs separated by $\sigma \pm \Delta \sigma / 2$ in the tangential plane and $\pi \pm \Delta \pi / 2$ along the line of sight. Note that in the absence of galaxy peculiar velocities $i)$ $\xi(s)$ would coincide with $\xi(r)$ and $ii)$ $\xi(\sigma,\pi)$ would contain no additional information compared to $\xi(s)$, since galaxy clustering is expected to be intrinsically isotropic. However, due to ``redshift-space distortions'' \citep{Kaiser1987}, the two-dimensional correlation function has a characteristic non-isotropic shape (see Fig.~\ref{fig:corr2d_hi}), and contains non-trivial cosmological information (see e.g. \citealp{Reid2012, Contreras2013}).

Lastly, we can measure the ``projected correlation function'', which is denoted by \Xis \ and is defined as\footnotemark{}

\begin{equation}
\Xi(\sigma) \, / \, \sigma = \frac{2}{\sigma} \int_{\pi=0}^{\pi=\pi_{max}} \xi(\sigma,\pi) \, d\pi \; \; ,
\end{equation}

\noindent
where $\pi_{max} = 45 \; h_{70}^{-1}$ Mpc is used in this article. \Xis \ is a correlation measure that is integrated over the line-of-sight direction. As a result, it is not affected by redshift space distortions and therefore is the most closely related to $\xi(r)$ . In fact, if the real space correlation function follows a power-law form, parametrized as $\xi(r) = (r\, / \, {r_0})^{-\gamma}$, then 

\begin{equation}
\Xi(\sigma)\,/\,\sigma = \frac{r_0^\gamma \: \Gamma(1/2) \Gamma((\gamma-1)/2)}{\Gamma(\gamma/2)} \;\; \sigma^{-\gamma} \;\; . 
\label{eqn:ksi_powerlaw}
\end{equation}

\noindent
In other words, a power-law projected correlation function has the same exponent $\gamma$ as the real space correlation function, while at the same time its normalization can be used to determine the clustering scale-length parameter, $r_0$. Table \ref{tab:powerlaw_fits} contains fitted values of $r_0$ and $\gamma$ parameters for all the projected correlation functions appearing in this article.

\footnotetext{There exists a different notation and definition for the projected correlation function that is more widely used in the literature, $w_p(r_p)$. In this case $r_p$ is the separation on the plane of the sky (same as $\sigma$) and $w_p = 2 \int_{\pi=0}^{\pi=\pi_{max}} \xi(r_p,\pi) \, d\pi$. Therefore, the two definitions differ only by a factor of $r_p \equiv \sigma$. In this article,  we opt for \Xis \ because $i)$ it is a unitless quantity and therefore independent of $H_0$ and $ii)$ has the same logarithmic slope as the real space correlation function, $\xi(r)$.}

 \subsection{Pair-weighting}
 \label{sec:pair_weighting}

In order to increase the effective volume probed by the ALFALFA and SDSS samples we weight each pair roughly inversely to the product of the individual selection function values for the two constituent objects. This weighting aims at taking into account the large number of pairs that remain undetected at large distances. More specifically, each data-data, random-random and data-random pair is counted towards $P_{DD}$, $P_{RR}$ and $P_{DR}$ with a weight $w_{ij}$ given by 

\begin{eqnarray}
w_{ij} &=& w_i \times  w_j \;,  \;\;\; \mathrm{where} \\
\label{eqn:weight} w_i &=& \frac{1}{1+4\pi\bar{n}J_3 \; \varphi(d_i)} \; .
\end{eqnarray}        

\noindent
In the expression above, $w_i$ and $w_j$ are the weights of object $i$ and $j$ respectively, while $\varphi(d_i)$ is the selection function at the distance of object $i$. $\bar{n}$ is the average volume-limited number density of the sample, while $J_3$ is a short hand notation for $J_3(s=30Mpc) = \int_{s =0}^{s=30\mathrm{Mpc}} \: s^2 \,  \xi(s) \: ds$. Results are not sensitive to the exact value of $J_3$, so a value of $J_3 = 2\,962 \: \mathrm{Mpc}^3$ is used here, corresponding to a fiducial $\xi_{fid}(s) = (s/5 \, \mathrm{Mpc})^{-1.5}$. In essence, the weight in Eqn.~\ref{eqn:weight} reduces to $w_{ij} \propto \frac{1}{\varphi_i} \times \frac{1}{\varphi_j}$ when the selection function is relatively large ($\varphi(d) \gg 1/(4\pi\bar{n}J_3)$), while $w_{ij} \approx 1$ when the selection function is small ($\varphi(d) \ll 1/(4\pi\bar{n}J_3)$).  


In the case of SDSS data-data pair counts, an additional weight is applied to correct for SDSS ``fiber collisions''. SDSS spectroscopic fibers cannot generally be placed closer than $55^{\prime\prime}$ from one another, lowering the counts of pairs at small on-sky separations. As a result, SDSS data-data pair weights are given by

\begin{equation}
w_{ij} = w_i \times w_j \times w_{fc}(\theta_{ij}) \; .
\end{equation}

\noindent
$w_i$ and $w_j$ are defined as per Eqn.~\ref{eqn:weight}, while $w_{fc}(\theta_{ij})$ is the fiber collision correction that depends only on the angular separation between the two galaxies. The analytic form of $w_{fc}(\theta)$ used in this article (Cheng Li, private communication) is the same as the one described in \citet{Li2006a} and tested with mock catalogs in \citet{Li2006b}. Note, however, that the separations probed in this article are relatively large ($\gtrsim 200$ kpc), and as a result the impact of fiber collisions on the projected correlation functions never exceeds the $\approx2\%$ level.


\subsection{Error estimation}   

We calculate errors on our clustering measurements by bootstrapping our data sample. If a data sample has $N_D$ elements, bootstrap resampling involves forming sample realizations by randomly extracting $N_D$ elements from the original data set, with replacement. If $k=1,\ldots, K$ sample realizations are produced in this way, we can calculate statistical properties of the measured correlations such as the average, variance and covariance matrix:

\begin{eqnarray}
\left< \xi_i \right> &=& \frac{1}{K} \; \sum_{k=1}^K \xi^{(k)}_i  \\
\sigma^2_{\xi_i} &=& \frac{1}{K-1} \; \sum_{k=1}^K (\xi^{(k)}_i - \left< \xi_i \right>)^2 \;\; \mathrm{and} \\
\mathrm{Cov}(\xi_i,\xi_j) &=& \frac{1}{K-1} \; \sum_{k=1}^K (\xi^{(k)}_i - \left< \xi_i \right>)(\xi^{(k)}_j - \left< \xi_j \right>) \; .
\end{eqnarray}

\noindent
Here, $\xi_i$ denotes generically the value of some correlation measure in separation bin $i$, while the superscript $^{(k)}$ denotes the $k^\mathrm{th}$ bootstrap sample realization. Note that, as is usual practice, we use in this article random catalogs with more objects than our data samples ($N_R > N_D$). This is done in order to ensure that the contribution of the counting noise of random-random pairs to the overall error budget is subdominant to the error from data-data and data-random pairs. Specifically, each random catalog is 10 times the length of the corresponding data catalog (up to a maximum of $100\,000$), and 25 bootstrap realizations are used to estimate the mean, variance and covariance. We note here that the number of bootstrap realizations is highly inadequate to compute accurate covariance matrices for the two-dimensional correlation functions presented in this article. As a result, we do not attempt to perform a rigorous quantitative analysis of these two-dimensional clustering measurements; we rather present them in order to offer better intuitive understanding of the results concerning the one-dimensional correlation functions. Finally, to ease the computational workload, the normalized random-random counts are only computed once, while the data-data and data-random pairs are computed for each realization.

\section{Results}
\label{sec:results}

\subsection{Dependence of clustering on HI mass}

\begin{figure}[htbp]
\centering
\includegraphics[scale=0.6]{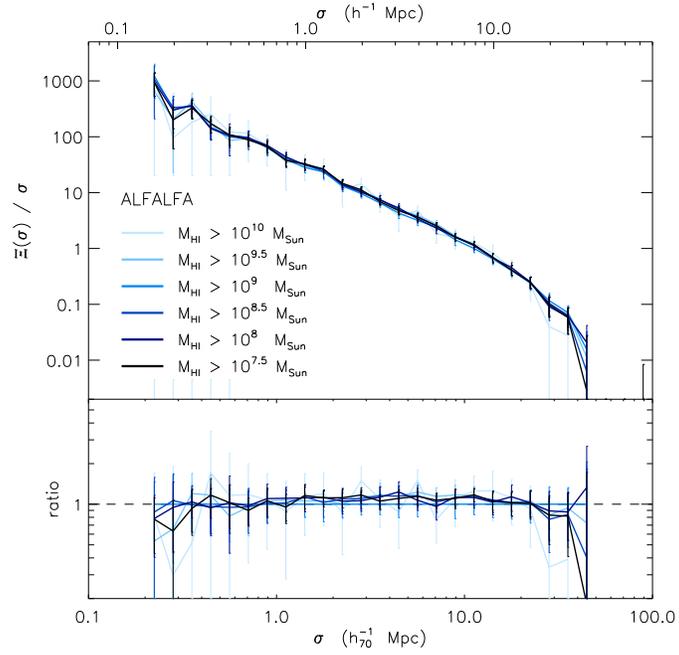}
\caption{ \footnotesize{\textit{upper panel:} The projected correlation function, \Xis, for the HI mass thresholded samples shown in the upper panel of Figure~\ref{fig:himass_hist} and described in \S\ref{sec:alfa_sample}. Darker shades of blue represent samples with a lower HI mass threshold, as depicted in the Figure legend. Note that some error bars for the $M_{HI} > 10^{10} \; M_\odot$ sample extend below the legend. \textit{lower panel:} The same projected correlation functions as above, normalized to the correlation function of the $M_{HI}> 10^9 \; M_\odot$ sample. The unity line is also plotted for reference.} }
\label{fig:corr_hithresh}
\end{figure}

Figure~\ref{fig:corr_hithresh} shows the measured projected correlation function, \Xis, for the HI mass thresholded samples shown in the upper panel of Figure~\ref{fig:himass_hist}, and described in \S\ref{sec:alfa_sample}. A few preliminary comments are worth making: Firstly, the projected correlation function for all samples is well approximated by a power-law, up to a length scale of $\sigma \approx 15 \; \mathrm{h}_{70}^{-1}\,\mathrm{Mpc}$. Secondly, the correlation functions seem to deviate from the simple power-law form at separations larger than this characteristic value. This behavior has been noted in multiple literature studies and seems to hold for both optically selected and HI-selected samples (e.g. \citealp{Li2012,Norberg2002,Zehavi2011} for optical samples  and \citealp{Martin2012, Basilakos2007} for HI samples).

\begin{figure}[htbp]
\centering
\includegraphics[scale=0.6]{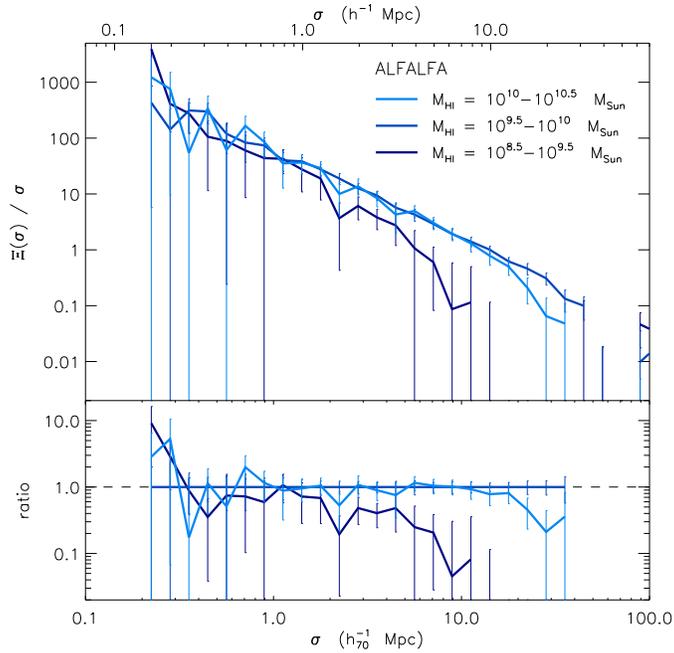}
\caption{ \footnotesize{\textit{upper panel:} The projected correlation function, \Xis, for the non-overlapping HI mass binned samples shown in the lower panel of Figure~\ref{fig:himass_hist} and described in \S\ref{sec:alfa_sample}. Darker shades of blue represent samples with a lower range of HI masses, as depicted in the Figure legend. \textit{lower panel:} The same projected correlation functions as above, normalized to the correlation function of the $M_{HI} = 10^{9.5}-10^{10} \; M_\odot$ sample. The unity line is also plotted for reference. Note that the range of the $y$-axis in the lower panel of this Figure is much larger than in Figure \ref{fig:corr_hithresh}.} }
\label{fig:corr_hibins}
\end{figure}

Most importantly however, the correlation functions of the HI mass thresholded samples show no significant differences among one another, within the errors of the present analysis. Note that Figure~\ref{fig:corr_hithresh} shows no evidence for enhanced clustering for the samples with the highest HI masses; this is in stark contrast to the strong clustering displayed by galaxies with high stellar mass or optical luminosity (e.g. \citealp{Zehavi2011,Beutler2013} to name a few, see also Fig.\ref{fig:corr_magthresh} in this work). Figure~\ref{fig:corr_hithresh} is not ideal however for assessing the clustering properties of low HI mass galaxies; as the upper panel of Figure~\ref{fig:himass_hist} puts in evedence, even the samples with the lowest HI mass thresholds (as low as $M_{HI} = 10^{7.5} \; M_\odot$) are still dominated by fairly HI massive galaxies. We therefore display in Figure~\ref{fig:corr_hibins} the \Xis \ measurements for the three non-overlapping HI mass binned samples shown in the lower panel of Figure~\ref{fig:himass_hist}. Galaxies with intermediate and high masses ($M_{HI} = 10^{9.5} - 10^{10} \; M_\odot$ and $M_{HI} = 10^{10} - 10^{10.5} \; M_\odot$ bins) show no significant differences in their clustering properties in Figure~\ref{fig:corr_hibins}, in accordance with the results in Figure~\ref{fig:corr_hithresh}. Interestingly enough though, galaxies with low HI mass ($M_{HI} = 10^{8.5} - 10^{9.5} \; M_\odot$ bin) seem to be much more weakly clustered than their more massive counterparts.

\begin{figure}[htbp]
\centering
\includegraphics[scale=0.6]{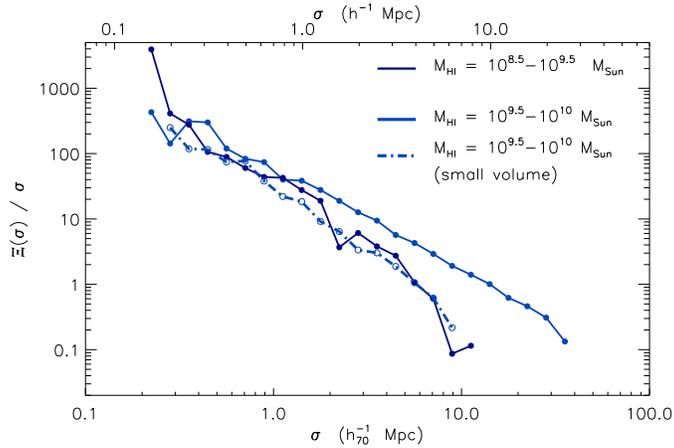}
\caption{ \footnotesize{The solid lines are the projected correlation functions for the $M_{HI} = 10^{8.5} - 10^{9.5} \; M_\odot$ and $M_{HI} = 10^{9.5} - 10^{10} \; M_\odot$ samples (darker and lighter shade, respectively). The lighter shade dash-dotted line shows again the correlation function of the $M_{HI} = 10^{9.5} - 10^{10} \; M_\odot$ sample, but this time restricted to the $\approx 6$ times smaller volume occupied by the galaxies in the $M_{HI} = 10^{8.5} - 10^{9.5} \; M_\odot$ sample. Error bars are omitted for clarity. This Figure shows that the difference in clustering between the two samples is probably due entirely to finite volume effects.} }
\label{fig:volume_effects}
\end{figure}

The HI mass dependence of the clustering properties of HI-selected galaxies has remained a controversial issue in the literature. For instance, \citet{Basilakos2007} and \citet{Meyer2007} have both analyzed datasets from the HIPASS survey, but came to different conclusions regarding the issue. On one hand, \citet{Basilakos2007} found that HIPASS galaxies with $M_{HI} < 10^{9.4} \; M_\odot$ have a significantly lower clustering amplitude than galaxies with HI masses larger than this value (see their Fig.~5). On the other hand, \citet{Meyer2007} dissected the HIPASS sample at a similar HI mass, $M_{HI} = 10^{9.25} \; M_\odot$, but found no convincing differences in the correlation function of the low-mass and high-mass subsamples (see their Fig.~12). At face value, the ALFALFA measurement shown in our Figure~\ref{fig:corr_hibins} seems to lend support to the \citet{Basilakos2007} claim. One complication arises however due to the fact that the volume probed by the $M_{HI} = 10^{8.5} - 10^{9.5} \; M_\odot$ sample is $\approx 6$ times smaller than the volume probed by the $M_{HI} = 10^{9.5} - 10^{10} \; M_\odot$ sample. As a result the observed discrepancy could be caused by finite volume effects. We therefore re-calculate the projected correlation function of the $M_{HI} = 10^{9.5} - 10^{10} \; M_\odot$ sample, but restricting it to the smaller volume available to the $M_{HI} = 10^{8.5} - 10^{9.5} \; M_\odot$ sample. Figure~\ref{fig:volume_effects} shows the result: Even though the correlation functions of the two samples are very different from one another when both are calculated over their full volumes (dark blue and blue solid lines), they show no significant differences when the two samples are restricted to a common volume (dark blue solid line and blue dash-dotted line).

Overall, we find no conclusive evidence for a dependence of the clustering properties of HI-selected galaxies on their HI mass, over the mass range $M_{HI} \approx 10^{8.5}-10^{10.5} \; M_\odot$. Despite the fact that Figure~\ref{fig:corr_hibins} displays a weak correlation function for low HI mass galaxies ($M_{HI} = 10^{8.5} - 10^{9.5} \; M_\odot$), Figure~\ref{fig:volume_effects} suggests that this behavior could be entirely due to finite volume effects. An extension of this work to both higher and lower masses will necessitate the next generation of HI surveys, such as the planned WALLABY survey with the ASKAP array \citep{Koribalski2012} and the HI surveys to be performed with the APERTIF instrument on the WSRT interferometer, which will probe a much larger volume than ALFALFA and are expected to detect $\approx$10x more sources. These surveys will also provide more accurate measurements of the correlation function over the HI mass range probed in this work, potentially uncovering trends that cannot be detected within current precision.


\subsection{Bias relative to optical galaxies}
\label{sec:rel_bias}

\begin{figure}[htbp]
\centering
\includegraphics[scale=0.6]{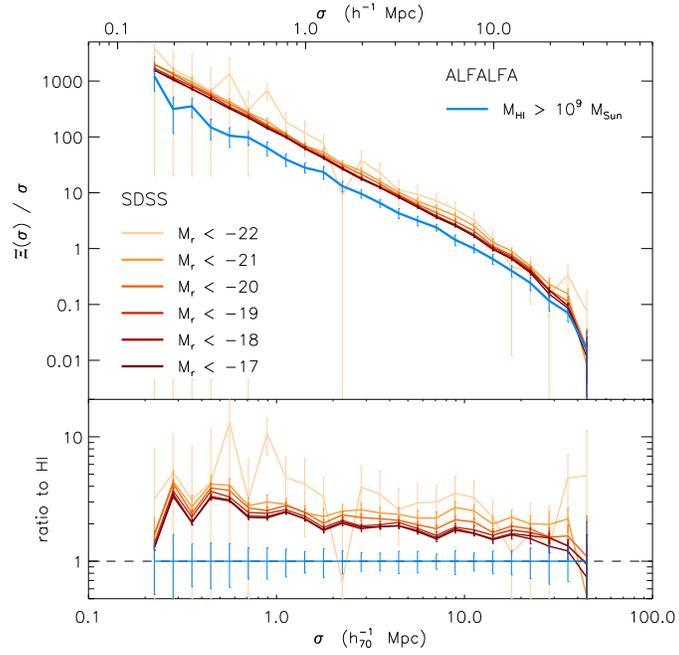}
\caption{ \footnotesize{\textit{upper panel:} The projected correlation function, \Xis, for the luminosity thresholded SDSS samples (shown in the upper panel of Figure~\ref{fig:mag_hist} and described in \S\ref{sec:sdss_sample}), compared to the correlation function of the $M_{HI} > 10^9 \; M_\odot$ ALFALFA sample. Darker shades of red represent samples with a lower optical luminosity threshold, as depicted in the Figure legend. Note that some of the error bars of the $M_r < -22$ sample extend below the legend. \textit{lower panel:} The same projected correlation functions as above, normalized to the correlation function of the ALFALFA sample. The unity line is also plotted for reference.} }
\label{fig:corr_magthresh}
\end{figure}

Several literature studies have found that HI-rich galaxies are among the most weakly clustered galactic populations known (e.g. \citealp{Basilakos2007, Meyer2007, Martin2012}). This fact can be clearly seen in Figure~\ref{fig:corr_magthresh}, which compares the projected correlation function of one representative  ALFALFA sample ($M_{HI}>10^9 \; M_\odot$) with the correlation functions of the luminosity thresholded SDSS samples (as depicted in the upper panel of Figure~\ref{fig:mag_hist} and described in \S\ref{sec:sdss_sample}). The Figure shows that a ``typical'' HI-selected sample is significantly less clustered than a ``typical'' optically selected sample, regardless of the optical sample's limiting luminosity.

\begin{figure}[htbp]
\centering
\includegraphics[scale=0.6]{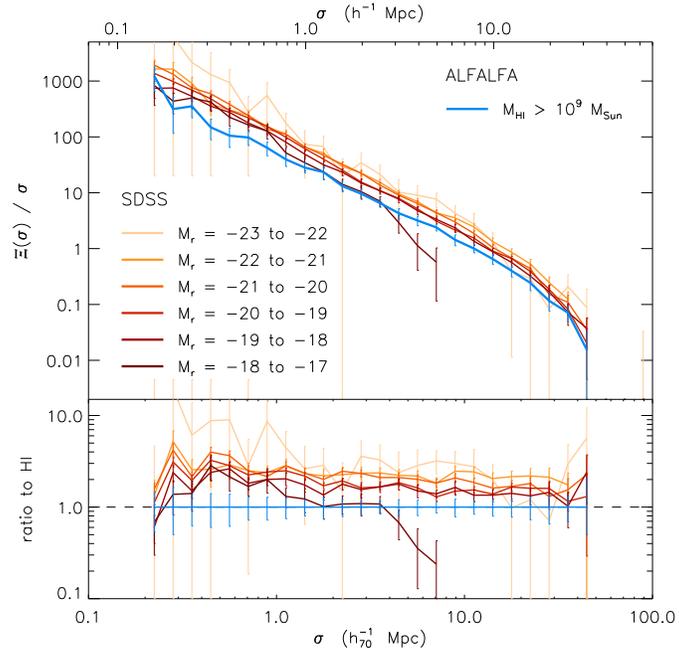}
\caption{ \footnotesize{The projected correlation function, \Xis, for the luminosity binned SDSS samples (shown in the lower panel of Figure~\ref{fig:mag_hist} and described in \S\ref{sec:sdss_sample}), compared to the correlation function of the $M_{HI} > 10^9 \; M_\odot$ ALFALFA sample. Darker shades of red represent samples with a lower range of optical luminosities, as depicted in the Figure legend. Note that some of the error bars for the $M_r =$ -23 -- -22 sample extend below the legend. \textit{lower panel:} The same projected correlation functions as above, normalized to the correlation function of the ALFALFA sample. The unity line is also plotted for reference.} }
\label{fig:corr_magbins}
\end{figure}

Furthermore, we compare the correlation function of the same ALFALFA sample to the correlation function of the SDSS luminosity-binned samples (as depicted in the lower panel of Figure~\ref{fig:mag_hist}). The result is shown in Figure~\ref{fig:corr_magbins}, on which we note the following points: Firstly, the optical samples display a clear trend  of stronger clustering with increasing luminosity, unlike HI mass-binned samples (Fig.~\ref{fig:corr_hibins}). The dependence of clustering on luminosity has been extensively studied in the literature, and the observed trend has been interpreted as a tendency of more luminous galaxies to inhabit more massive DM halos (see e.g. \citealp{Zehavi2011, Beutler2013} for two recent examples). Secondly, and most importantly, the correlation function of the HI-selected sample is lower in amplitude than the correlation function of even relatively faint optical galaxies (at least as faint as $M_r \approx -18$). In addition,  the optically selected samples seem to display a steeper correlation function regardless of luminosity.\footnotemark{}

\footnotetext{Note that the $M_r = -18 \; \mathrm{to}  -17$ sample has not been taken into account when making the statements above. The correlation function of this sample is probably affected significantly by finite-volume effects, similarly to the case of the $M_{HI} = 10^{8.5}-10^{9.5} \; M_\odot$ ALFALFA sample (see Fig.~\ref{fig:volume_effects}).}

\begin{figure}[htbp]
\centering
\includegraphics[scale=0.6]{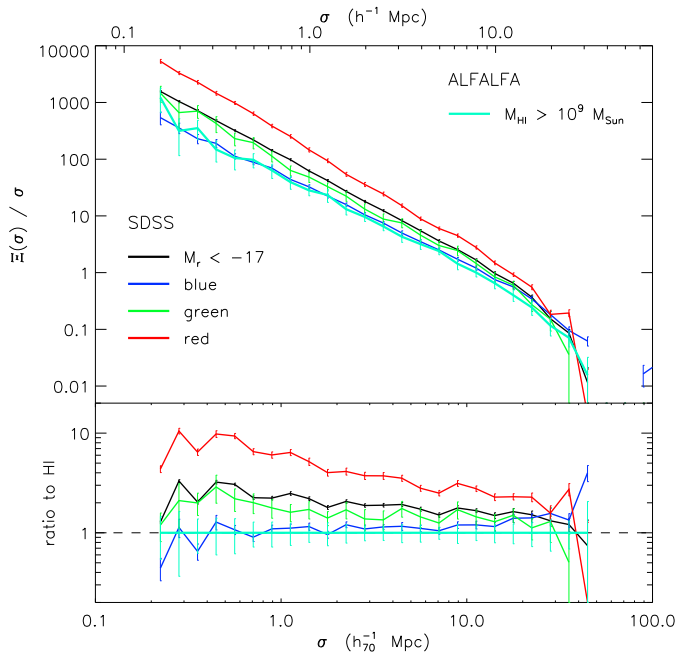}
\caption{ \footnotesize{\textit{upper panel:}The projected correlation function, \Xis, for the ``blue'', ``green'' \& ``red'' SDSS samples (shown in Figure~\ref{fig:cmd} and described in \S\ref{sec:sdss_sample}), compared to the correlation function of the $M_{HI} > 10^9 \; M_\odot$ ALFALFA sample. \textit{lower panel:} The same projected correlation functions as above, normalized to the correlation function of the ALFALFA sample. The unity line is also plotted for reference. } }
\label{fig:corr_colors}
\end{figure}

Figure~ \ref{fig:corr_colors}, on the other hand, compares the clustering of the same $M_{HI} > 10^9 \; M_\odot$ ALFALFA sample to the clustering of three optical subsamples split by color (see Fig.~\ref{fig:cmd} and \S\ref{sec:sdss_sample}). By combining the information in Figs. \ref{fig:corr_magthresh}, \ref{fig:corr_magbins} \& \ref{fig:corr_colors}, we arrive at the following conclusions:

\begin{enumerate}


\item HI-selected galaxies cluster less than optically selected galaxies, when no color cuts are applied to the latter sample. This statement is valid even for relatively faint galaxies (at least as faint as $M_r \approx -18$). In addition, optically selected samples of all luminosities display slightly steeper correlation functions compared to HI-selected samples. 



\item The correlation function of HI-selected galaxies is practically indistinguishable from the correlation function of optical galaxies with blue colors. The relative bias\footnotemark{} of the two samples is $b_{rel} \approx 1$, over almost the whole range of separations probed. 


\item Red galaxies show much stronger clustering than HI-selected galaxies, with the relative bias reaching values $b_{rel} > 3$ at small separations ($\sigma \lesssim 1$ Mpc). Moreover, the projected correlation function of red optical galaxies is significantly steeper than that of HI-selected galaxies.

\end{enumerate}


\footnotetext{The relative bias between two samples $s_1$ \& $s_2$ is defined as the square root of the ratio of their real space correlation functions, in other words $b^2_{rel}(r) \equiv \xi_{s_1}(r)/\xi_{s_2}(r)$. Bias values quoted in this article are calculated by fitting the projected correlation function according to Eqn. \ref{eqn:ksi_powerlaw}, under the assumption of a power-law form for the real space correlation function, $\xi(r)$ (see Table \ref{tab:powerlaw_fits}).}

\begin{figure}[htbp]
\centering
\includegraphics[scale=0.6]{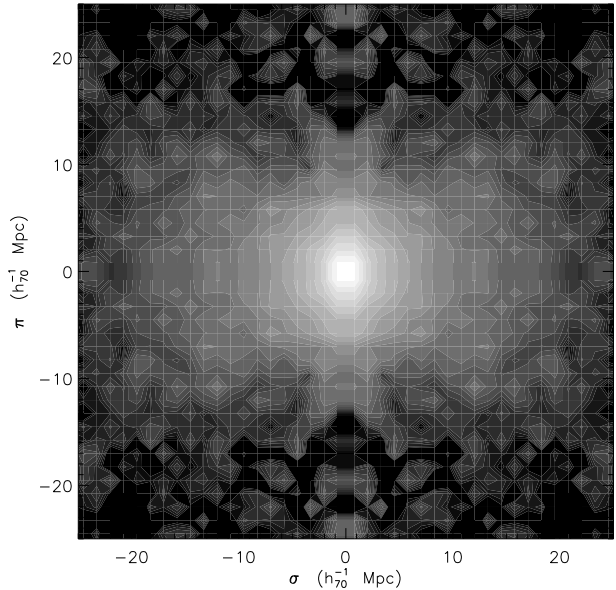}
\caption{ \footnotesize{The two-dimensional correlation function, \xisp, of the ALFALFA parent sample ($M_{HI} > 10^{7.5} \; M_\odot$). Note that \xisp \ is calculated in \textit{linear} bins of separation, with  $\sigma_{min} = \pi_{min} = 0.15  \; h_{70}^{-1}\,\mathrm{Mpc}$ and bin size $\Delta\sigma=\Delta\pi=1.25 \; h_{70}^{-1}\,\mathrm{Mpc}$. The contours are logarithmically spaced, starting at a value of 0.05 and increasing by a factor of 2 every three contours up to a factor of 6.3. Note also that all the information of \xisp \ is contained in one quadrant of the plot;  the other three quadrants are just mirrored copies.} }
\label{fig:corr2d_hi}
\end{figure}

\begin{figure}[htbp]
\centering
\includegraphics[scale=0.6]{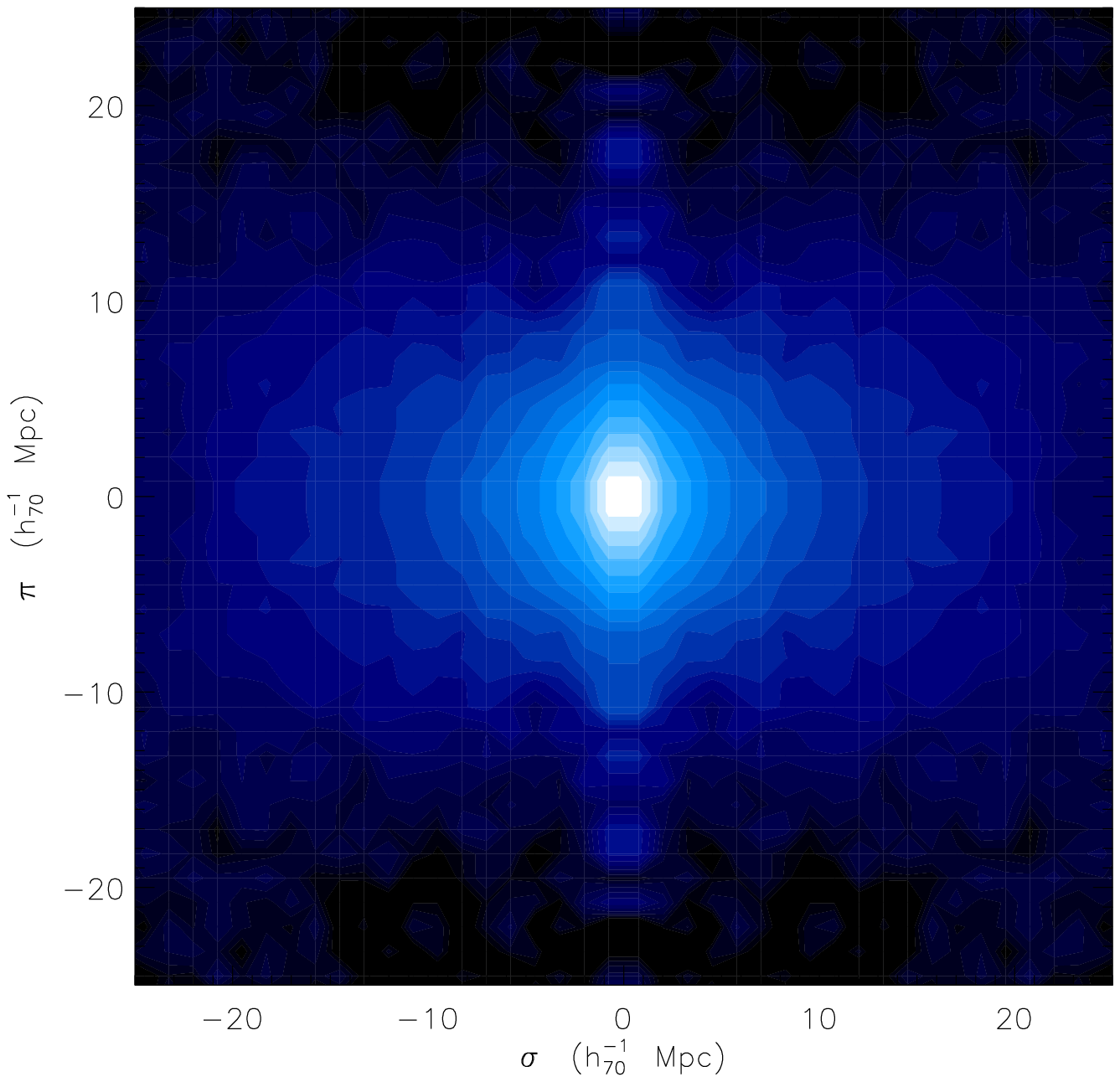}
\caption{ \footnotesize{ The two-dimensional correlation function, \xisp, of the SDSS ``blue'' sample (see Fig. \ref{fig:cmd}). The separation bins and contour levels are the same as for Fig. \ref{fig:corr2d_hi}. } }
\label{fig:corr2d_blue}
\end{figure}

\begin{figure}[htbp]
\centering
\includegraphics[scale=0.6]{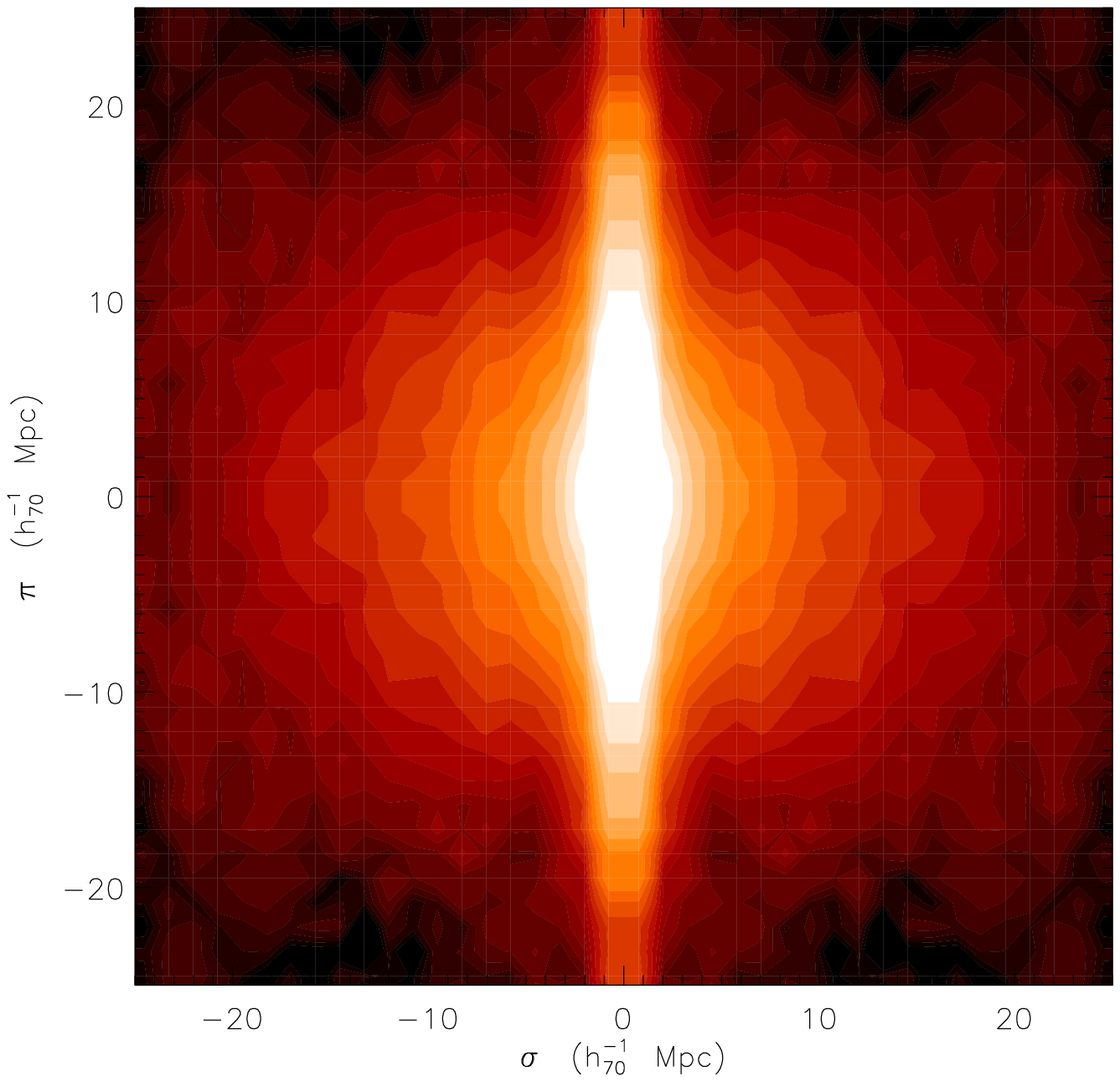}
\caption{ \footnotesize{ The two-dimensional correlation function, \xisp, of the SDSS ``red'' sample (see Fig. \ref{fig:cmd}). The separation bins and contour levels are the same as for Fig. \ref{fig:corr2d_hi}.  } }
\label{fig:corr2d_red}
\end{figure}

Points 1-3 above hold also for the full two-dimensional correlation function, $\xi(\sigma,\pi)$: Figure~\ref{fig:corr2d_hi} shows $\xi(\sigma,\pi)$ for the parent ALFALFA sample ($M_{HI}>10^{7.5} \; M_\odot$), which can be compared with $\xi(\sigma,\pi)$ for the blue and red SDSS subsamples (Figures~\ref{fig:corr2d_blue} \& \ref{fig:corr2d_red}, respectively). Note that common contour levels are used in Figs. \ref{fig:corr2d_hi} - \ref{fig:corr2d_red}. The two-dimensional correlation functions for the ALFALFA and blue SDSS galaxies are very similar in amplitude and shape. In particular, both samples display a characteristic ``flattening'' of $\xi(\sigma,\pi)$ along the $\pi$-axis on intermediate scales ($\pi \gtrsim 10$ Mpc), as well as a weak ``finger of god'' effect (i.e. the elongated structure along the $\pi$-axis at $\sigma \approx 0$ Mpc). By contrast, the red SDSS subsample shows a $\xi(\sigma,\pi)$ with much larger overall amplitude, as well as a very distinct finger of god feature. In addition, the $\xi(\sigma,\pi)$ contours for the red SDSS sample display a more symmetric, ``round'', shape on intermediate scales ($\gtrsim 10$ Mpc). We remind the reader that a rigorous, quantitative analysis of the two-dimensional correlation functions in Figures \ref{fig:corr2d_hi}-\ref{fig:corr2d_red} is beyond the scope of this article; they are presented here in order to provide some intuitive understanding of the results observed in the one-dimensional correlation functions.

These results are not unexpected; it is well established that gas-rich galaxies are associated with late-type morphology, blue optical colors and elevated specific star formation rates (e.g. \citealp{Huang2012b, Catinella2010, Li2012b}). For example, \citet{Huang2012b} shows that the ALFALFA sample is heavily biased against red-sequence galaxies, while sampling very well the less luminous and more actively star-forming galaxies galaxies in the ``blue cloud'' (their Fig. 11). The main conclusions summarized in points 1-3, therefore, are a direct consequence of the fact that blue galaxies are significantly less clustered than red galaxies, irrespective of luminosity (see e.g. Fig 16 in \citealp{Zehavi2011}). The bias of blind HI surveys against red-sequence galaxies also helps explain the marked difference in the shape of $\xi(\sigma,\pi)$ between the ALFALFA and SDSS red samples (Figs. \ref{fig:corr2d_hi} \& \ref{fig:corr2d_red}, respectively). Red galaxies are usually found in high density environments, such as clusters of galaxies and compact groups, and their clustering bears the signs of large and incoherent peculiar motions which are characteristic of these environments. In particular, the red sample has an increased number of galaxy pairs that have small physical but large velocity separations; these pairs produce the strong ``finger of god'' feature in \xisp \ at $\sigma \approx 0$. On the other hand, galaxies with blue colors and HI galaxies tend to inhabit  the lower density ``field''. As a result, they trace the ordered flow towards matter overdensities without significant noise from peculiar motions. This  is why the characteristic asymmetric shape of \xisp \ at separations $\gtrsim$10 Mpc, which is caused by these systematic motions, is more pronounced in the blue and HI samples.

\subsection{Cross-correlation between HI-selected and optically selected samples}


The results above can be used to compare the clustering properties of HI and optical galaxies, but do not contain information about the spatial relationship among the samples under consideration. In particular, they cannot address questions such as whether or not HI galaxies inhabit the same environments as a given class of optical galaxies. It is already known through the study of individual clusters \citep{Giovanelli1985, Haynes1986, Solanes2002} that galaxies in high density environments tend to have lower gas fractions than their counterparts in the field, and thus have a lower probability of being included in an HI-selected sample. Since optical galaxies with red colors are found preferentially in dense environments, we expect HI-selected samples to show some degree of ``segregation'' with respect to red galaxies. Here, we use the large galaxy samples provided by the ALFALFA and SDSS surveys to obtain a statistical measurement of this effect, and to pin down the length scale over which environment can affect the gaseous contents of galaxies.    


\begin{figure}[htbp]
\centering
\includegraphics[scale=0.6]{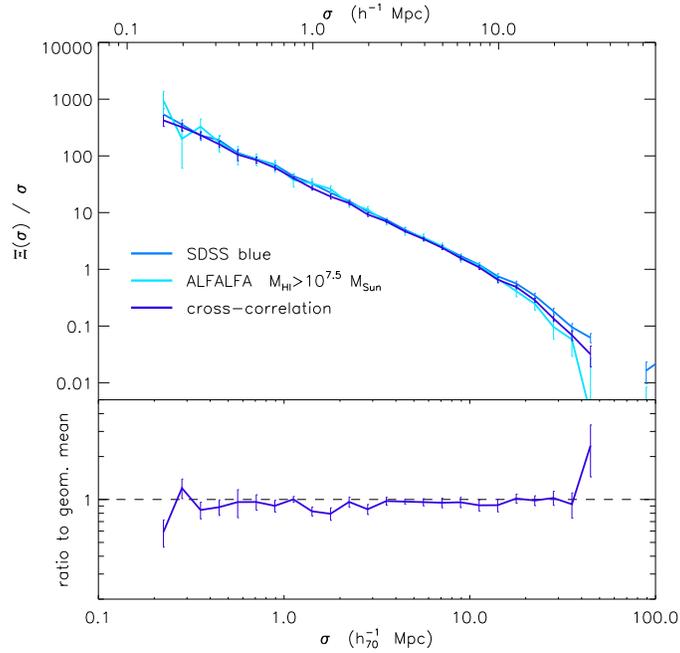}
\caption{ \footnotesize{ \textit{upper panel:} The projected correlation function, \Xis, for the blue SDSS sample (dark blue solid line) and the parent ALFALFA sample (light blue solid line), compared to their projected cross-correlation function (purple solid line). \textit{lower panel:} The ratio of the cross-correlation function to the geometric mean of the correlation functions of the two constituent samples, $\mathcal{R}(\sigma)$. } }
\label{fig:xcorr_hi_blue}
\end{figure}

\begin{figure}[htbp]
\centering
\includegraphics[scale=0.6]{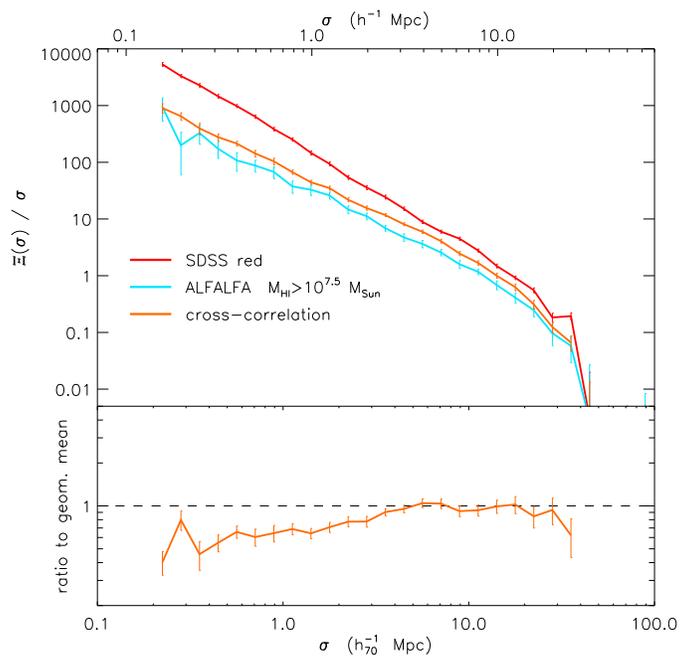}
\caption{ \footnotesize{  \textit{upper panel:} The projected correlation function, \Xis, for the red SDSS sample (red solid line) and the parent ALFALFA sample (light blue solid line), compared to their projected cross-correlation function (orange solid line). \textit{lower panel:} The ratio of the cross-correlation function to the geometric mean of the correlation functions of the two constituent samples, $\mathcal{R}(\sigma)$. Note the clear tendency for SDSS galaxies with red colors and HI-rich ALFALFA galaxies to avoid each other at separations $\lesssim 3$ Mpc.} }
\label{fig:xcorr_hi_red}
\end{figure}

The spatial relationship of two galactic samples is encoded in their cross-correlation function. In this article, we calculate cross-correlation functions using a modified version of the LS estimator (following \citealp{Zehavi2011}): 

\begin{equation}
\hat{\xi}_{cross} = (DD_1DD_2 -DD_1RR_2 - DD_2 RR_1 + RR_1RR_2 )/RR_1RR_2 \;\; .
\end{equation}  

\noindent 
Here $DD$, $RR$ and $DR$ are the normalized data-data, random-random and data-random pair counts, and the subscripts 1 \& 2 are used to denote the two samples. Generally, the information present in the cross-correlation function is most intuitively presented in terms of its ratio with the geometric mean of the correlation functions of the two constituent samples, $\mathcal{R}(r) \equiv\xi_{cross}(r)/\sqrt{\xi_1(r) \xi_2(r)}$. In essence, $\mathcal{R}$ measures the degree to which two samples are spatially ``aware'' of one another: Two spatially independent samples have $\mathcal{R}=1$ for all $r$ (i.e. $\xi_{cross}(r) = \sqrt{\xi_1(r) \xi_2(r)}$). Conversely, a ratio of $\mathcal{R}< 1$ at some separation $r$ means that the two samples ``avoid'' each other on the length scale under consideration.


Figure~\ref{fig:xcorr_hi_blue} shows the cross-correlation function between the ALFALFA parent sample ($M_{HI} > 10^{7.5} \; M_\odot$) and the blue SDSS sample. The lower panel shows that $\mathcal{R} \approx 1$ on all probed scales, meaning that HI-selected galaxies and optical galaxies with blue colors have no special spatial relationship. In other words, detecting a blue SDSS galaxy at a given location in space does not influence our chance of finding an ALFALFA-detected galaxy in its vicinity, beyond what is expected from the clustering of the samples. The situation is very different in Figure~\ref{fig:xcorr_hi_red}, which shows that  the cross-correlation function between HI-selected galaxies and optical galaxies with red colors is systematically lower than their geometric mean at small separations (i.e. $\mathcal{R} < 1$ at $\sigma \lesssim 3$ Mpc). This means that the existence of a red SDSS galaxy at a given position in space lowers the chances that an HI-rich galaxy is positioned within $\approx 3$ Mpc from it. 

\begin{figure}[htbp]
\centering
\includegraphics[scale=0.6]{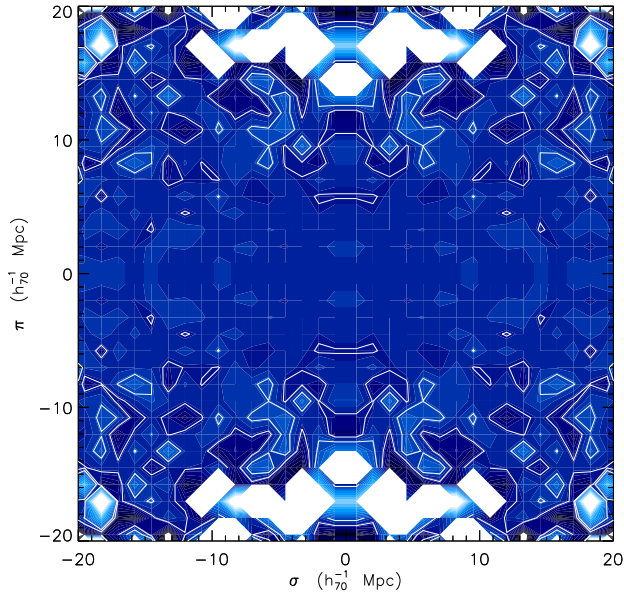}
\caption{ \footnotesize{Two-dimensional map of $\mathcal{R}(\sigma,\pi)$ for the ALFALFA parent sample ($M_{HI} > 10^{7.5} \; M_\odot$) and the SDSS blue sample. $\mathcal{R}(\sigma,\pi)$ is the ratio of the cross-correlation function between the two samples to the geometric mean of their respective correlation functions. The separation bins are the same as in Fig. \ref{fig:corr2d_hi}. The contour levels are logarithmically spaced, with values doubling every six contours. The darkest shade corresponds to the minimum value of 0.25 while the lightest shade corresponds to the maximum value of 4. Regions enclosed by solid contours are regions where $\mathcal{R}$ deviates significantly from unity ($\mathcal{R} < 0.85$ or $\mathcal{R} > 1.15$).  } }
\label{fig:xcorr2d_hi_blue}
\end{figure}

\begin{figure}[htbp]
\centering
\includegraphics[scale=0.6]{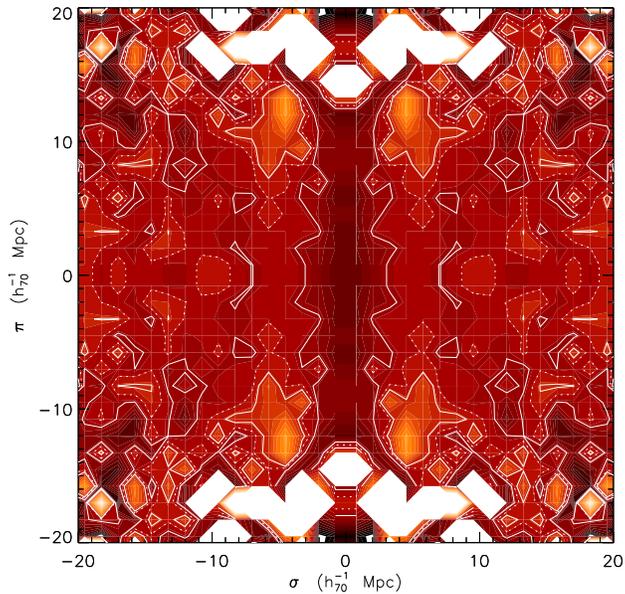}
\caption{ \footnotesize{Same as Fig.~\ref{fig:xcorr2d_hi_blue} but regarding the cross-correlation of the ALFALFA parent sample ($M_{HI} > 10^{7.5} \; M_\odot$) with the SDSS red sample. The separation bins and contour levels are also the same as in Fig.~\ref{fig:xcorr2d_hi_blue}. Regions enclosed by solid contours are regions where $\mathcal{R}$ deviates significantly from unity ($\mathcal{R} < 0.85$ or $\mathcal{R} > 1.15$). Since the central part of the map contains systematically low values of $\mathcal{R}$, an additional dotted contour level at $\mathcal{R} = 1$ has been drawn to lift ambiguities. } }
\label{fig:xcorr2d_hi_red}
\end{figure}

These results also hold for the two-dimensional cross-correlation functions between the HI-selected ALFALFA sample and the color-based SDSS samples. Figure~\ref{fig:xcorr2d_hi_blue}, for example, shows a two-dimensional map of $\mathcal{R}(\sigma,\pi)$ calculated from the cross-correlation function between the ALFALFA parent sample and the SDSS blue sample. Regions that are enclosed by solid contours are those where $\mathcal{R}$ deviates significantly from unity ($\mathcal{R}<0.85$ or $\mathcal{R} > 1.15$). We can clearly see that, barring the large fluctuations at the outskirts of the map caused by noise, $\mathcal{R}(\sigma,\pi) \approx 1$ over most of the map. The situation is  very different when the cross-correlation between the ALFALFA parent sample and the SDSS red sample is considered. Figure \ref{fig:xcorr2d_hi_red} shows that regions corresponding to $\sigma \lesssim 3$ Mpc have systematically low values of $\mathcal{R}$, over the whole range of $\pi$-axis separations. This characteristic shape demonstrates graphically that HI-selected galaxies avoid  regions of space where the finger-of-god effect is large, corresponding mostly to galaxy clusters and rich groups. Once again, we note that in this article we do not aim to perform a quantitative analysis of the two-dimensional distributions in Figures \ref{fig:xcorr2d_hi_blue} \& \ref{fig:xcorr2d_hi_red}. We present them in order to offer some intuition regarding the results of the one-dimensional cross-correlation analysis.


In general, measurements of the cross-correlation properties of HI galaxies with respect to various optical samples are especially important in the context of cosmological studies with next generation HI surveys \citep[e.g.][]{Beutler2011}, and 21cm intensity-mapping experiments at moderate redshift \citep[e.g.][]{Masui2013}. In particular, Figs.~\ref{fig:corr_magthresh} - \ref{fig:xcorr2d_hi_red} show that an HI-selected sample traces the cosmic large-scale structure differently than most optical surveys. For example, due to the very different clustering properties of HI-rich galaxies and galaxies with red colors,   a 21cm survey would provide a very different view of the large scale structure compared to a survey of, e.g., luminous red galaxies (as in e.g. \citealp{Eisenstein2005}). On the other hand, a survey targeting actively star-forming galaxies (such as the UV-selected WiggleZ survey; \citealp{Drinkwater2010}) will be a much closer match in terms of clustering properties and spatial distribution. The considerations above have an effect on the potential of future 21cm surveys for cosmological studies: For example, the low clustering amplitude of HI galaxies may lower our sensitivity for detecting the BAO feature in the large-scale galaxy correlation function. On the other hand, measurements that are based on the anisotropy of \xisp \ may benefit considerably from the low levels of peculiar motion ``noise'' achieved with an HI-selected sample. In addition, the low bias of HI samples leads to smaller deviations from linearity compared to other highly biased samples, potentially aiding the cosmological interpretation of clustering measurements. As a result, HI surveys may prove advantageous in measuring redshift-space distortions (RSD) and the growth of structure (``$f\sigma_8$'' measurements, e.g. \citealp{Reid2012, delaTorre2013, Contreras2013}).

\section{Which halos host gas-rich galaxies?}  
\label{sec:halos}  

\subsection{\LCDM \ halo sample}
\label{sec:bolshoi_sample}

We select a sample of dark matter halos from the Bolshoi \LCDM \ simulation \citep{Klypin2011}. The Bolshoi simulation is a high-resolution dissipationless simulation, run for a set of cosmological parameters consistent with the 7-year results of the \textit{Wilkinson Microwave Anisotropy Probe} \citep[WMAP;][]{Jarosik2011} and other recent cosmological studies. 
We use the halo catalogs\footnotemark{} extracted with the Bound-Density-Maxima (BDM) halo finder algorithm \citep{KlypinHoltzman1997}.  
A detailed description of the BDM halo finding method can be found in Appendix A of \citet{Klypin2011}. Here we briefly summarize its main characteristics: the algorithm identifies density peaks in the survey volume, and associates neighboring particles into halos. The virial radius of each halo is defined as the radius enclosing an average overdensity of about 360 times the cosmic matter density (at $z=0$). Unbound particles are removed iteratively, and a number of properties are then recorded for each halo (e.g. virial mass, maximum circular velocity, angular momentum, etc.). In the case of subhalos (i.e. density maxima laying within the virial radius of a larger halo), the extent is confined to particles that are bound to the substructure. The Bolshoi simulation is one of the highest resolution cosmological simulations available today, and is complete down to a maximum circular velocity of 50 \kms \ (or $M_{vir} \approx 2 \cdot 10^{10} \; h_{70}^{-1}\,M_\odot$).

\begin{figure}[htbp]
\centering
\includegraphics[scale=0.7]{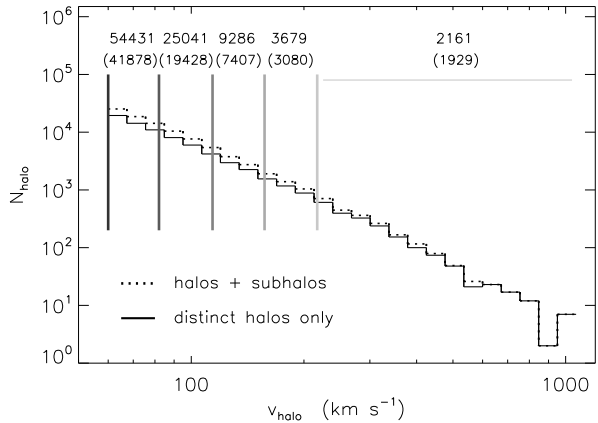}
\caption{ \footnotesize{Histogram of maximum circular velocities, $v_{halo}$, for halos selected from the Bolshoi \LCDM \ simulation \citep{Klypin2011}. The solid histogram represents the counts of distinct halos only, while the dotted histogram refers to all halos (including both distinct halos as well as subhalos). The vertical solid lines denote the boundaries of the five velocity-binned samples, described in \S\ref{sec:bolshoi_sample}. The quoted numbers correspond to the overall halo count in each sample (halos \& subhalos), while the numbers in parentheses denote the number of distinct halos only. Note that, unlike the similar Figs.~\ref{fig:himass_hist} \& \ref{fig:mag_hist}, both axes in this Figure are logarithmic.  } }
\label{fig:vhalo_hist}
\end{figure}

In particular, we select a box region of the Bolshoi simulation of size $\approx$$140 \; h_{70}^{-1}$Mpc on a side, such that the volume of the halo sample is comparable to the ALFALFA volume. In addition, we restrict ourselves to halos with maximum circular velocities $v_{halo} > 60$ \kms, in order to exclude halos that are close to the resolution limit of the simulation. In total, our halo sample consists of $94\,671$ halos, including both distinct halos as well as subhalos. From this parent halo sample we create subsamples by specifying five $v_{halo}$ ranges, as shown in Figure~\ref{fig:vhalo_hist}: 60-82 \kms, 82-114 \kms, 114-157 \kms, 157-217 \kms \ and $>$217 \kms.

\footnotetext{\texttt{www.multidark.org/MultiDark/Help?page=databases/bolshoi/database}}


Furthermore, we consider halo subsamples split by their spin parameter,  $\lambda_K$, defined as:

\begin{equation}
 \lambda_K = \frac{J\, K^{1/2}}{G\,M_{vir}^{5/2}} \;\; .
 \label{eqn:spin}
 \end{equation}

\begin{figure}[htbp]
\centering
\includegraphics[scale=0.7]{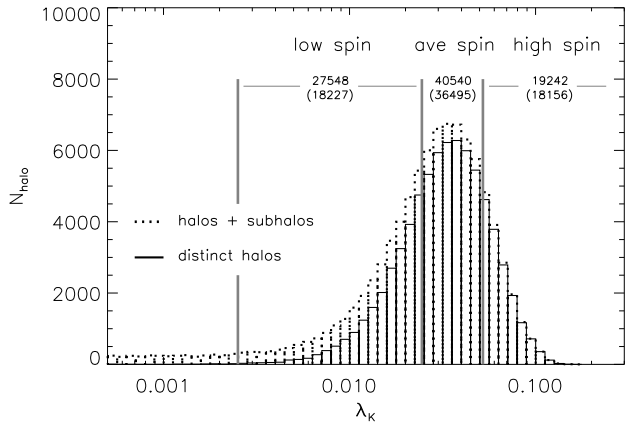}
\caption{ \footnotesize{Histogram of ``kinetic'' spin parameter, $\lambda_K$ (see Eqn.\ref{eqn:spin} for definition), for halos selected from the Bolshoi \LCDM \ simulation \citep{Klypin2011}. The solid histogram represents the counts of distinct halos only, while the dotted histogram refers to all halos (including both distinct halos as well as subhalos). The vertical solid lines denote  the boundaries of the three spin-based samples. The quoted numbers correspond to the overall halo count in each sample (halos \& subhalos), while the numbers in parentheses denote the number of distinct halos only. } }
\label{fig:spin_hist}
\end{figure}

\noindent 
In the Equation above, $J$ is the total angular momentum of the particles within the virial radius of the halo, $K$ is the total kinetic energy of these particles (not including the energy associated with the bulk motion of the halo through space), and $M_{vir}$ is the halo virial mass. Note that the BDM database for the Bolshoi simulation reports spin parameters defined in terms of the halo kinetic energy ($K$), instead of the more common definition based on total energy ($\lambda = J|E_{tot}|^{1/2} / GM_{vir}^{5/2}$). However, the two definitions yield very similar results for well-virialized halos, since in this case $K \approx |E_{tot}|$. Figure \ref{fig:spin_hist} displays graphically the three spin-based subsamples, referred to as ``low spin'', ``average spin'' and ``high spin'' samples.

The halo samples described above are volume-limited, meaning that inclusion in some specific sample does not depend on their position in the simulation box. As a result, all halo samples share the same random catalog, which is straightforwardly created by a set of points with uniformly distributed $x,y,z$ coordinates. In addition, no pair-weighting (see \S\ref{sec:pair_weighting}) is necessary, since all halo pairs in the simulation can be accounted for. Lastly, the separations between pairs of halos are readily available in terms of physical length, and are not affected by redshift-space distortions. It follows that for these halo samples $\xi(r) \equiv \xi(s)$, while \Xis \ can be calculated by projecting separations on any arbitrary axis.

\subsection{Halo mass \& halo/subhalo status}

\begin{figure}[htbp]
\centering
\includegraphics[scale=0.6]{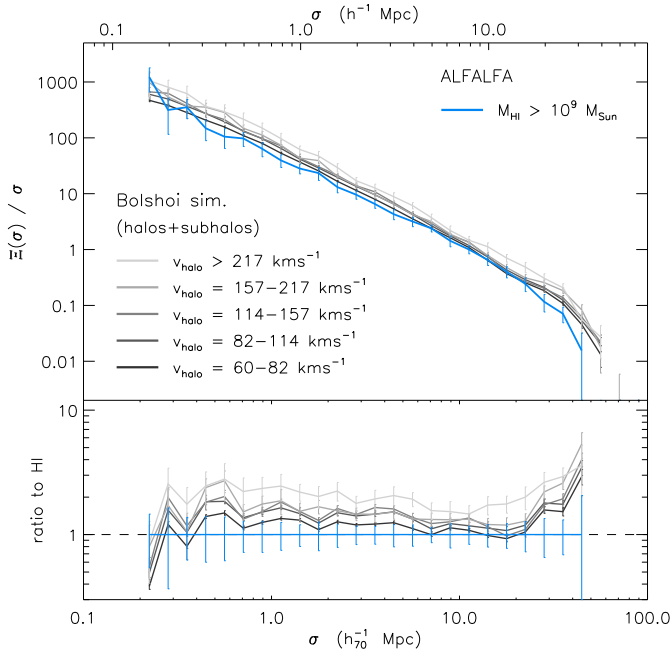}
\caption{ \footnotesize{  \textit{upper panel:} The projected correlation function for the velocity-binned Bolshoi halo samples (shown in Figure~\ref{fig:vhalo_hist} and described in \S\ref{sec:bolshoi_sample}), compared to the correlation function of the $M_{HI} > 10^9 \; M_\odot$ ALFALFA sample. Both halos \& subhalos are included in the computation of the halo \Xis. Darker shades represent samples a lower $v_{halo}$ range, as depicted in the Figure legend. \textit{lower panel:} The same projected correlation functions as above, normalized to the correlation function of the ALFALFA sample. The unity line is also plotted for reference.   } }
\label{fig:corr_velbins_hs}
\end{figure}

\begin{figure}[htbp]
\centering
\includegraphics[scale=0.6]{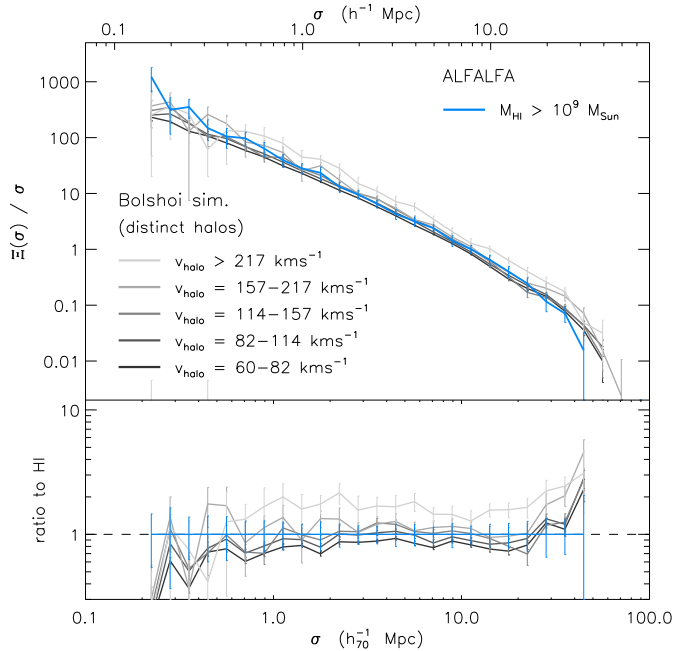}
\caption{ \footnotesize{ Same as in Fig.~\ref{fig:corr_velbins_hs}, but for distinct halos only. Note that some error bars for the $v_{halo} > 217$ \kms \ sample extend below the legend. } }
\label{fig:corr_velbins_dh}
\end{figure}

Figures \ref{fig:corr_velbins_hs} \& \ref{fig:corr_velbins_dh} show the projected correlation functions of the velocity-binned halo samples, including and excluding subhalos from the computation respectively.
Overplotted on both Figures is the projected correlation function of the $M_{HI}>10^9 \; M_\odot$ ALFALFA sample (solid cyan line). In both Figures there is a clear trend for more massive halos to show increasingly stronger clustering. This trend is consistent with theoretical expectations \citep[e.g.][]{Musso2012}, since more massive halos are expected to form in regions with larger matter overdensity. 
This behavior is not shared by the HI mass-thresholded samples in this work, which do not display any discernible clustering enhancement with increasing HI mass (see Figure \ref{fig:corr_hithresh}). This fact alone suggests that galaxy HI mass may not be tightly related to the mass of the host halo.  
%
Furthermore, a comparison of Figures \ref{fig:corr_velbins_hs} \& \ref{fig:corr_velbins_dh} shows that the inclusion of subhalos in a sample leads to higher amplitude clustering, especially at small separations. This is also expected, since subhalos are found in the vicinity of other halos by definition. 
More specifically Figure \ref{fig:corr_velbins_hs} shows that, when subhalos are included, all halo samples display stronger clustering than ALFALFA galaxies. This further suggests that a sizable fraction of subhalos do not host HI galaxies. According to Figures  \ref{fig:corr_velbins_hs} \& \ref{fig:corr_velbins_dh}, we expect ALFALFA galaxies with $M_{HI}>10^9 \; M_\odot$ to be hosted by halos with $v_{halo}\sim$~80-150~\kms, depending on the fraction of subhalos hosting HI galaxies.

Alternatively, we can study the relation between our HI-selected sample and their host dark matter halos more systematically, by using the technique of abundance matching. Abundance matching is a simple, yet powerful statistical approach to connect galaxy properties (such as the luminosity, stellar or baryonic mass, velocity, etc.) to their host dark matter (sub)halos \citep[e.g.,][]{Shankar2006,Conroy2006,Guo2010,Reddick2012,Rodriguez2012,Papastergis2012}. In its most simple form, the observed abundances of galaxies at a given property are matched against the theoretical halo plus subhalo abundances. The result is a galaxy property versus halo mass empirical relation. In reality, galaxy properties are not determined solely by the mass of the halo in which they reside but, due the complexity of the galaxy formation process, a dependence on other halo and/or environmental properties is expected. To take this into account, recent works have extended the abundance matching technique to include a scatter around the mean relation \citep[e.g.,][]{Behroozi2010,Moster2010,Hearin2012,Rodriguez2013}.

\begin{figure}[htbp]
\centering
\includegraphics[scale=0.6]{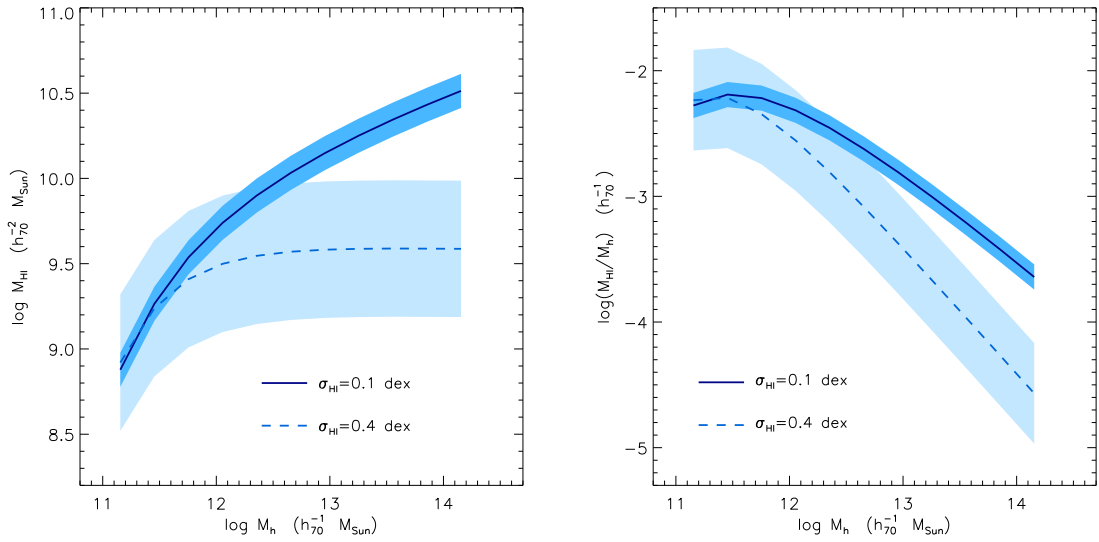}
\caption{ \footnotesize{\emph{Left panel:} The solid and dashed lines represent the $M_{HI}$-$M_h$ relation obtained via abundance matching, assuming a scatter of $\sigma_{HI}=0.1$ dex and $\sigma_{HI}=0.4$ dex, respectively. The shaded areas denote the assumed scatter around each average relation. Note that, in the $\sigma_{HI}=0.4$ dex case, the HI mass is nearly independent of halo mass for $M_{HI} \gtrsim 10^{9.5} \; M_\odot$.
\emph{Right panel:} The solid and dashed curves represent the $M_{HI} / M_h$ ratio as a function of $M_h$ in the $\sigma_{HI}=0.1$ dex and $\sigma_{HI}=0.4$ dex cases, respectively. Note that in both cases, the maximum $M_{HI}/M_h$ value is reached at $M_h \approx 10^{11.2}\; M_\odot$, which is significantly lower than the halo mass where the maximum of the $M_\ast/M_h$ ratio occurs ($M_h \approx 10^{12}\; M_\odot$).
} }
\label{fig:mh1mh}
\end{figure}



Here we use the abundance matching technique in order to obtain an average relation between galaxy HI mass and host (sub)halo mass ($M_{HI}$-$M_h$ relation). To do so, we employ the observed HI mass function of the HI-selected sample of \citet[][see their Table 1]{Papastergis2012} and the halo plus subhalo mass function obtained from the Bolshoi simulation (see \S \ref{sec:bolshoi_sample}). \citet{Reddick2012} have shown that the measure of halo mass that is most tightly related to the stellar properties of galaxies (e.g. stellar mass or optical luminosity) is the maximum mass reached along the entire merger history of the (sub)halo, $M_{h,{\rm peak}}$; we therefore perform our abundance matching analysis using $M_{h,{\rm peak}}$ as the halo mass, dropping from now on the excess notation ($M_{h,{\rm peak}} \rightarrow M_h$). Note that $M_{h,{\rm peak}}$ is approximately equal to the present-day mass for distinct halos, but in the case of subhalos it can be significantly larger than their present-day mass, due to the effects of tidal stripping. 

We furthermore assume that the distribution of HI mass at a given (sub)halo mass is drawn from a lognormal distribution with mean $M_{HI}(M_h)$ and a scatter of $\sigma_{HI}$ around it. Here we will assume that $\sigma_{HI}$ is independent of halo mass. While the scatter around the average stellar mass-halo mass relation ($M_*(M_h)$ relation) has been discussed extensively in the literature \citep[e.g.,][]{Cacciato2009,More2009,Yang2009,More2011,Leauthaud2012,Rodriguez2013}, $\sigma_{HI}$ has not been discussed previously. Here we opt to use two different values for $\sigma_{HI}$ in order to gauge the uncertainty introduced from our lack of knowledge on its value and mass dependence: $\sigma_{HI}=0.1$ dex and $\sigma_{HI}=0.4$ dex. A more thorough exploration of this scatter is deferred for a future publication.


\begin{figure}[htbp]
\centering
\includegraphics[scale=0.6]{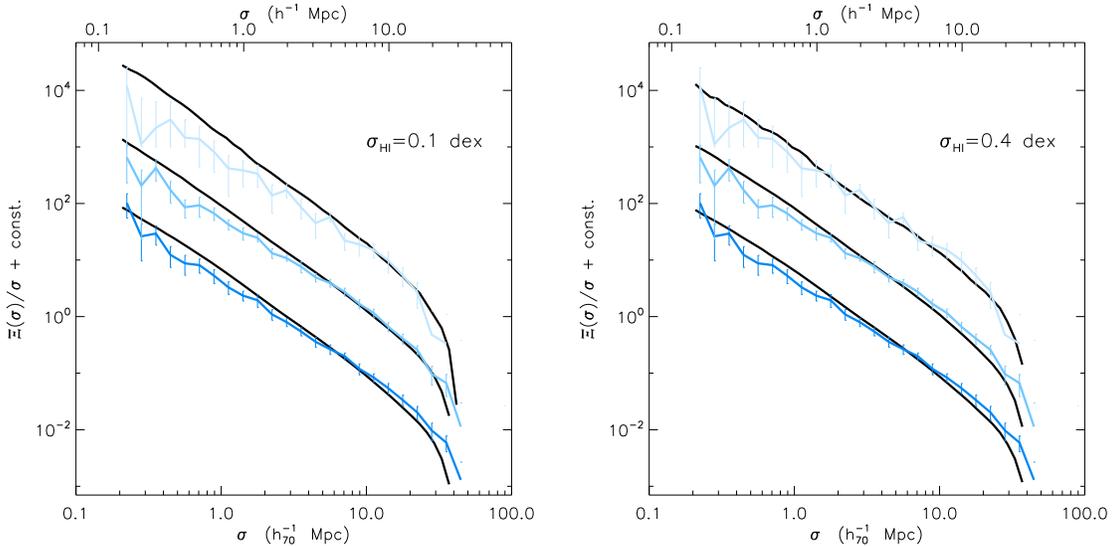}
\caption{ \footnotesize{Comparison of the correlation functions of the ALFALFA HI mass-thresholded samples with simulated samples obtained by assigning 
$M_{HI}$ values to each Bolshoi (sub)halo according to the relation in Fig. \ref{fig:mh1mh}. Each ALFALFA sample is plotted in a different shade of blue, while all simulated samples are plotted as black solid lines. From top to bottom each pair of lines corresponds to the $M_{HI} > 10^{10},10^{9.5}, 10^9 \; M_\odot$ samples (a constant offset between samples has been used for clarity). Note that the lower ends of some error bars for the $M_{HI} > 10^{10} \; M_\odot$ ALFALFA sample have been clipped for clarity. \textit{Left panel:} $M_{HI}$ values are assigned to each Bolshoi (sub)halo via abundance matching assuming $\sigma_{HI}=0.1$ dex (solid line in Fig. \ref{fig:mh1mh}).  \textit{Right panel:} Same as left panel, but assuming $\sigma_{HI}=0.4$ dex (dashed line in Fig. \ref{fig:mh1mh}). Observations seem to favor a larger value of scatter.}  }
\label{fig:2pcf_modelHI}
\end{figure}

The left-hand panel of Figure~\ref{fig:mh1mh} shows the resulting average $M_{HI}(M_h)$ relations for both values of $\sigma_{HI}$. Note that, while for the $\sigma_{HI} = 0.1$ dex case the $M_{HI}(M_h)$ relation is clearly monotonic at all masses, in the $\sigma_{HI}=0.4$ dex case $M_{HI}$ is nearly independent of halo mass (except perhaps for $M_{HI} \lesssim 10^9 \; M_\odot$). According to the abundance matching result, galaxies with $M_{HI} > 10^9 \; M_\odot$ are hosted by halos with $M_h \gtrsim 10^{11.3} \; M_\odot$, or $v_{halo} \gtrsim 100$ \kms, in reasonable agreement with our claims based on Figures \ref{fig:corr_velbins_hs} \& \ref{fig:corr_velbins_dh}. The right-hand panel of Figure~\ref{fig:mh1mh} shows the average $M_{HI}/M_h$ ratio as a function of $M_h$. The maximum of the HI-to-halo mass ratio is obtained at $M_h\approx10^{11.3}M_{\odot}$ in the case of $\sigma_{HI}=0.1$ dex, and at $M_h\approx10^{11.1}M_{\odot}$ in the $\sigma_{HI}=0.4$ dex case. Note that both values are lower than the values commonly obtained for the location of the peak of the $M_\ast/M_h$ ratio, which is $M_h\approx10^{12}M_{\odot}$.




Once we have assigned a value of $M_{HI}$ to each (sub)halo of the Bolshoi simulation, we can compute the correlation functions of modeled samples with any range of HI mass, and compare them with the ALFALFA results. For example, in Figure \ref{fig:2pcf_modelHI} we compare the projected correlation functions of three ALFALFA HI mass-thresholded samples to the correlation functions of modeled samples with the same HI thresholds. 
Note that, due to the $v_{halo} > 60$ \kms \ cut that we have imposed on the Bolshoi halos, our parent halo sample is complete only down to $M_h \approx 10^{11} \; M_\odot$; in order to avoid biases related to this selection criterion, we conservatively restrict our abundance matching analysis to galaxies with $M_{HI} > 10^9 \; M_\odot$.
Overall, we find that the clustering dependence on HI mass is rather weak. Nevertheless, in the $\sigma_{HI}=0.1$ dex case modeled galaxies with large HI masses ($M_{HI}>10^{10}M_{\odot}$) show stronger clustering than their lower HI mass counterparts. On the other hand, in the $\sigma_{HI}=0.4$ dex case the clustering amplitude of modeled galaxies is almost independent of HI mass. This latter case is therefore in better agreement with the observational results of Fig. \ref{fig:corr_hithresh}. We conclude that the clustering properties of ALFALFA galaxies favor an $M_{HI}$-$M_h$ relation with a large scatter and a very weak dependence on halo mass (at least for galaxies with $M_{HI} \gtrsim 10^{9.5} \; M_\odot$). 

Moreover, Figure \ref{fig:2pcf_modelHI} shows that all modeled samples display consistently stronger clustering than the actual ALFALFA samples. This is because our abundance matching analysis assumes that all subhalos host an HI galaxy. If we repeat the analysis by considering only distinct halos, we find the opposite result: all modeled samples consistently underestimate the clustering amplitude of the actual ALFALFA samples (Figure not shown). Our second conclusion is therefore that only a subset of subhalos host HI galaxies, with the rest presumably hosting gas-poor galaxies that are not detected by ALFALFA. In view of the results above, it is important to ask whether halo properties other than mass and halo/subhalo status may be playing a major role in determining the gas content of galaxies.

\subsection{Halo spin parameter}

The spin parameter of the host halo has been suggested to be the decisive factor in setting a number of galaxy properties, such as the the galaxy's stellar and gas surface density. In fact, low surface brightness (LSB) galaxies are currently believed to be hosted by halos with higher-than-average spin parameters \citep[e.g.][]{BoissierPrantzos2000}. Several lines of evidence also suggest that halo spin may be closely related to the overall gas-to-stellar mass ratio of a galaxy, in the sense that halos with higher spin parameters host more gas-rich systems at fixed stellar mass. Firstly, gas-rich galaxies are known to be of relatively low surface brightness and low stellar mass density \citep{Catinella2010, Zhang2009,Huang2012b}, properties that are typically associated with high spin halos. \citet{Huang2012b} have furthermore directly estimated the galactic spin parameter of the entire ensemble of ALFALFA galaxies, obtaining the result that measured galaxy spin increases with increasing gas-fraction (their Figure 14).  Lastly, recent hydrodynamical simulations by \citet{Kim&Lee2012} have shown that, at fixed halo mass, halos with higher spin parameters have more extended gaseous disks and larger overall gas-to-stellar mass ratios (Ji-hoon Kim, private communication).      


\begin{figure}[htbp]
\centering
\includegraphics[scale=0.6]{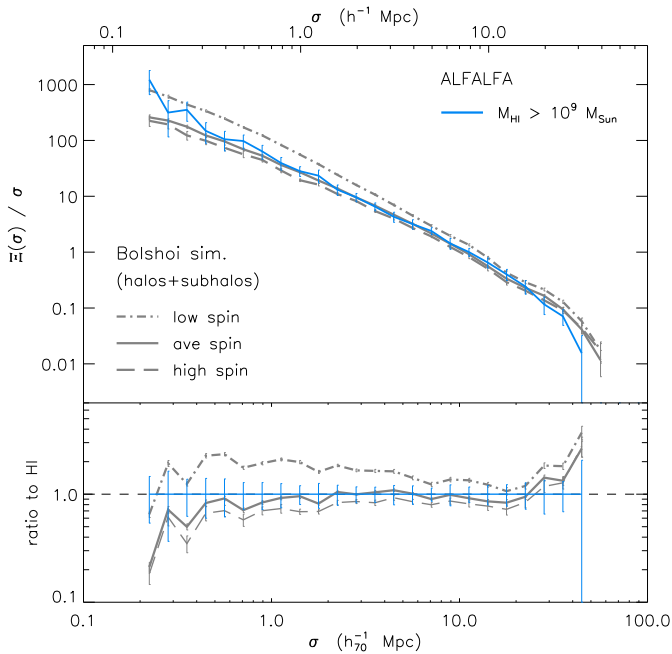}
\caption{ \footnotesize{ \textit{upper panel:} The projected correlation function for the Bolshoi halo samples split by spin (shown in Figure~\ref{fig:spin_hist} and described in \S\ref{sec:bolshoi_sample}), compared to the correlation function of the $M_{HI} > 10^9 \; M_\odot$ ALFALFA sample (cyan line). Both halos \& subhalos are included in the computation of the halo \Xis. The solid, dot-dashed and long-dashed grey lines represent the correlation functions of the ``average spin'', ``low spin'' and ``high spin'' samples, respectively. \textit{lower panel:} The same projected correlation functions as above, normalized to the correlation function of the ALFALFA sample. The unity line is also plotted for reference.   } }
\label{fig:corr_spinbins}
\end{figure}

Here we aim to investigate the spin-gas content relation by comparing the clustering properties of the gas-rich ALFALFA galaxies and of halos with ``low'', ``average'' and ``high'' spin parameters. Figure \ref{fig:corr_spinbins} shows the projected correlation functions for the three halo samples, and compares them with the projected correlation function of the $M_{HI}>10^9 \; M_\odot$ ALFALFA sample (solid cyan line). Note that both distinct halos and subhalos are included in the computation of \Xis \ for these halo samples, and no cuts are performed based on halo velocity. The Figure clearly shows that halos with low spin parameters display a markedly stronger correlation function compared to halos with average and high spin parameters. This behavior can be attributed to two main causes: Firstly, the ``low'' spin sample contains proportionally more subhalos than the ``average'' and ``high'' spin samples (see Fig.\ref{fig:spin_hist}), which tend to be a highly clustered population. Secondly, low spin parameter halos tend to cluster significantly more than their higher spin counterparts, even when the sample is restricted to distinct halos only (plot not shown). Note that the results hold also when spins are measured according to the ``classical'' definition, $\lambda = J\,|E_{\mathrm{tot}}|^{1/2} \, / \, G \, M_h^{5/2}$. 

We caution however the reader that the results presented above are sensitive to the algorithm used to identify halos and extract their properties. We have carried out an alternative analysis using a halo catalog for the Bolshoi simulation extracted by the ROCKSTAR\footnotemark{} halo finder \citep{Behroozi2013a, Behroozi2013b}. In contrast with BDM halos, ROCKSTAR halos do \textit{not} display clustering variations when split based on their spin parameter. At the same time, ROCKSTAR halos have a different distribution of spin parameters than BDM halos, with the most conspicuous difference being the lack of the low-spin tail in the distribution of ROCKSTAR subhalos (cf. Fig. \ref{fig:spin_hist}). The difference in the spin measurements between the two halo finders seems to be caused by the fact that BDM measurements are made in spherical shells, while ROCKSTAR measurements can be performed over significantly asymmetric domains, both in real and velocity space (Anatoly Klypin, private communication). In this article we choose to focus on results obtained with the BDM halo finder. The BDM algorithm has been used extensive in the literature and several aspects of the performance of the algorithm have been studied in detail. 


\footnotetext{The ROCKSTAR halo finding algorithm is a phase-space (6D) friends-of-friends (FOF) algorithm with an adaptively refined linking length. The method is described in detail in \citet{Behroozi2013a}, but here we give a basic overview of its operation: firstly, the survey volume is split into large groups by a position-space (3D) FOF algorithm. Then a phase-space linking length is calculated based on the position and velocity dispersions of the particles in the group. Structures are identified via a 6D FOF algorithm with the predefined linking length. The process above is repeated in the newly identified structures, to detect substructure. When no more substructure can be identified, unbound particles are excluded, and the structure properties of all substructures are measured. Catalogs of halos extracted from the Bolshoi simulation with the ROCKSTAR algorithm can be found at \texttt{www.slac.stanford.edu/$\sim$behroozi/Bolshoi\_Catalogs/} .   }

Aside from a potential sensitivity to the details of halo finding, the result above is remarkable, since it shows that a crude selection of halos based solely on their spin parameter can reproduce fairly well the clustering of ALFALFA galaxies. This finding lends further support to the hypothesis that halo spin --and perhaps not halo mass-- is the main property setting the gas content of galaxies. The result could also have important implications for the modeling of gas-rich galaxies, which will be an integral part of the scientific interpretation of near-future 21cm large-scale surveys. In particular, it may be necessary for semi-analytic models (SAMs) of HI galaxies \citep[e.g.][]{Obreschkow2009, Lu2012} to consider halo spin --in addition to mass-- in their implementation. Empirical approaches such as abundance matching may also need to be revised, and may potentially need to consider matching HI galaxy abundances with the joint distribution of halo mass and spin (see e.g. \citealp{HearinWatson2013} for a similar approach pertaining to galaxy optical colors).

\section{Conclusions}
\label{sec:conclusion}


\begin{deluxetable}{ccccc}

\tabletypesize{\footnotesize}
\tablecolumns{5}
\tablewidth{\linewidth}
\tablecaption{Power-law fits to all projected correlation functions.}

\tablehead{

\colhead{sample} &
\colhead{$N_{gal}$} &
\colhead{$\bar{n}$} &
\colhead{$r_0$} &
\colhead{$\gamma$} \\
\colhead{    }  &
\colhead{    } &
\colhead{($h_{70}^{3}\,\mathrm{Mpc}^{-3}$)} &
\colhead{($h_{70}^{-1}\,$Mpc)} &
\colhead{    }   
}

\startdata

\cutinhead{HI-mass thresholded samples}

7.5    &    $6\,123$   &    0.05562   &  4.74$\, \pm \,$0.20     &    1.604$\, \pm \,$0.049   
\\
8.0    &      $5\,980$  &  0.03561  &  4.85$\, \pm \,$0.20  &    1.642$\, \pm \,$0.055   
\\
8.5    &    $5\,650$  &   0.02147  &   4.75$\, \pm \,$0.24  &   1.613$\, \pm \,$0.058   
\\
9.0    &   $5\,160$  &   0.01123   &  4.53$\, \pm \,$0.23    &  1.624$\, \pm \,$0.062    
\\
9.5    &     $3\,789$  &  0.004135   &   4.77$\, \pm \,$0.22  &     1.607$\, \pm \,$0.058  
\\
10.0  &  $1\,211$  &  0.0007632  &  4.83$\, \pm \,$0.54  &     1.606$\, \pm \,$0.129  
\\

\cutinhead{HI-mass binned samples}

8.5 -- 9.5    &   $1\,861$   &   0.01733  &   3.49$\, \pm \,$0.31     &    2.190$\, \pm \,$0.266 
\\
9.5 -- 10    &   $2\,578$  &   0.003372   &  5.12$\, \pm \,$0.36  &    1.554$\, \pm \,$0.069   
\\
10 -- 10.5    &  $1\,181$  &  0.0007377  &   5.14$\, \pm \,$0.49  &   1.598$\, \pm \,$0.109   
\\

\cutinhead{magnitude thresholded samples}

-17    &  $18\,516$ &  0.02029  &   6.925$\, \pm \,$0.046     &    1.772$\, \pm \,$0.013   
\\
-18    &  $17\,025$  &  0.01288  &      6.945$\, \pm \,$0.045  &    1.775$\, \pm \,$0.012   
\\
-19    &   $12\,914$  &  0.007952   &     7.144$\, \pm \,$0.061  &   1.781$\, \pm \,$0.017  
\\
-20    & $7\,026$  &   0.004287  &   7.505$\, \pm \,$0.073    &  1.759$\, \pm \,$0.019    
\\
-21    &   $2\,571$  &  0.001576    &    8.011$\, \pm \,$0.168  &     1.701$\, \pm \,$0.039  
\\
-22    & $386$  &   0.0002320  &   9.311$\, \pm \,$0.661  &     1.816$\, \pm \,$0.145  
\\

\cutinhead{magnitude binned samples}

-17 -- -18  &  $1\,491$  &  0.007412  &   4.347$\, \pm \,$0.196     &    2.366$\, \pm \,$0.120  
\\
-18 -- -19   &  $4\,111$  &   0.004928  &   6.297$\, \pm \,$0.148  &    1.738$\, \pm \,$0.042   
\\
-19 -- -20   &   $5\,888$  &   0.003664  &     6.694$\, \pm \,$0.081  &    1.826$\, \pm \,$0.027   
\\
-20 -- -21    &  $4\,455$   &  0.002711  &    7.394$\, \pm \,$0.122  &   1.752$\, \pm \,$0.032   
\\
-21 -- -22    &  $2\,185$  &  0.001344  &  7.651$\, \pm \,$0.207    &  1.682$\, \pm \,$0.047  
\\
-22 -- -23    &     $381$   &   0.0002291  &  8.847$\, \pm \,$0.756  &     1.739$\, \pm \,$0.162  
\\
\\
\\

\cutinhead{optical color samples}

blue    &   $9\,031$  & 0.009896  &  4.767$\, \pm \,$0.094     &    1.589$\, \pm \,$0.025   
\\
green   &    $2\,622$  & 0.002873  &    6.061$\, \pm \,$0.204  &    1.719$\, \pm \,$0.057   
\\
red   &   $6\,863$  &  0.007520  &     9.892$\, \pm \,$0.104  &    1.982$\, \pm \,$0.021   
\\

\cutinhead{velocity binned halo samples (halos \& subhalos)}

60 -- 82    &   $53\,858$  &  0.01847  &  5.142$\, \pm \,$0.028     &    1.701$\, \pm \,$0.009   
\\
82 -- 114   &   $24\,769$  &  0.008494  &   5.668$\, \pm \,$0.039  &    1.720$\, \pm \,$0.013   
\\
114 -- 157   &   $9\,190$ &  0.003151  &    5.874$\, \pm \,$0.068  &    1.715$\, \pm \,$0.021   
\\
157 -- 217    &  $3\,634$  &  0.001246  &   5.952$\, \pm \,$0.134  &   1.746$\, \pm \,$0.038   
\\
$>$217    &   $2\,143$ &    0.0007349   &   6.982$\, \pm \,$0.194    &  1.780$\, \pm \,$0.055    
\\

\cutinhead{velocity binned halo samples (distinct halos)}

60 -- 82    &  $41\,459$ &  0.01421  &  3.848$\, \pm \,$0.043     &    1.567$\, \pm \,$0.011   
\\
82 -- 114   &    $19\,213$ &  0.006588    &   4.274$\, \pm \,$0.066  &    1.615$\, \pm \,$0.017   
\\
114 -- 157   &  $7\,332$  & 0.002513  &      4.535$\, \pm \,$0.110  &    1.605$\, \pm \,$0.026   
\\
157 -- 217    & $3\,041$  & 0.001042  &   5.083$\, \pm \,$0.229  &   1.668$\, \pm \,$0.058   
\\
$>$217    & $1\,912$  &  0.0006553  &   6.340$\, \pm \,$0.257    &  1.708$\, \pm \,$0.058    
\\

\cutinhead{spin parameter halo samples (halos \& subhalos)}

low    &    $27\,548$ & 0.009442  &   6.318$\, \pm \,$0.033     &    1.791$\, \pm \,$0.011   
\\
ave   &   $40\,540$  &   0.01389   &      4.355$\, \pm \,$0.034  &    1.605$\, \pm \,$0.009   
\\
high   &  $19\,242$  &   0.006595   &      3.744$\, \pm \,$0.075  &    1.551$\, \pm \,$0.020   
\\

\enddata

\label{tab:powerlaw_fits}

\tablecomments{The fits reported above assume a power-law real space correlation function, of the form $\xi(r) = (r/r_0)^{-\gamma}$. The parameters are derived by fitting the measured projected correlation functions according to Equation \ref{eqn:ksi_powerlaw}, over the separations $\sigma = 0.7$ -- $14 \;\; h_{70}^{-1}\,\mathrm{Mpc}$. The parameters and their errors were calculated with \texttt{mpfitfun} least-square fitting procedure, written in the IDL programming language.}

\end{deluxetable}

We use the sample of galaxies detected by the ALFALFA blind 21cm survey, to study the clustering characteristics of HI-selected galaxies (i.e. galaxies selected based on their atomic hydrogen content). In particular, we divide the ALFALFA galaxies into subsamples based on their HI mass, creating six HI mass-thresholded and three HI mass-binned samples, spanning the entire $M_{HI} = 10^{7.5} - 10^{11} \; M_\odot$ range. We measure the projected correlation function for each of the samples above, and find no compelling evidence for a dependence of clustering on HI mass. The data does yield a lower amplitude correlation function for the least massive HI mass-binned sample ($M_{HI} = 10^{8.5} - 10^{9.5} \; M_\odot$), but we attribute this effect to the small volume probed by the specific sample (see Figures \ref{fig:corr_hibins} \& \ref{fig:volume_effects}).

Moreover, we compare the clustering characteristics of the HI sample with those of optically selected galaxies drawn from the SDSS spectroscopic database. We follow a similar procedure as described above, and divide the optical galaxies into subsamples based on thresholds and ranges on their $r$-band absolute magnitude (see Figure \ref{fig:mag_hist}). In addition, we split the parent optical sample into three color-based subsamples, based on the galaxies' position on a color-magnitude diagram (see Figure \ref{fig:cmd}). We find that HI-selected galaxies cluster more weakly than their optical counterparts, even those at faint absolute magnitudes (at least as faint as $M_r \approx -18$). On the other hand, we find that the correlation function of ALFALFA galaxies is matched extremely well by the correlation function of SDSS galaxies with blue colors. Conversely, SDSS galaxies with red colors display much stronger clustering than the HI-selected samples, resulting in a projected correlation function with a markedly steeper slope and higher amplitude. The results above hold also for the full two-dimensional correlation function, \xisp: both the ALFALFA and SDSS blue samples display a strongly anisotropic shape at scales $\gtrsim$10 Mpc, and a very weak ``finger of god'' feature at small on-sky separations (see Figures \ref{fig:corr2d_hi} \& \ref{fig:corr2d_blue}). On the other hand, the two-dimensional correlation function of SDSS red galaxies shows a prominent finger of god feature and a more isotropic shape at intermediate scales (Figure \ref{fig:corr2d_red}).

In addition, we carry out a cross-correlation analysis between the ALFALFA and color-based SDSS samples. The HI$\times$red cross-correlation function shows that the gas-rich ALFALFA galaxies ``avoid'' being located within $\approx$3 Mpc of optical galaxies with red colors. In particular, they avoid environments where the finger of god effect is strong, presumably corresponding to clusters and rich groups. This amounts to a statistical measurement of the ``HI deficiency'' of galaxies in clusters, and yields a quantitative result for the length scale over which dense environments typically affect the gas contents of galaxies. 

We also measure the clustering properties of the halos in the Bolshoi \LCDM \ simulation, to gain insights on the characteristics of halos hosting gas-rich galaxies. By comparing the clustering of ALFALFA galaxies to that of halo samples split based on their mass and halo/subhalo status we arrive at the conclusions that $i$) HI mass is not tightly related to the mass of the host halo and $ii$) a sizable fraction of subhalos does not host gas-rich galaxies. We furthermore perform a more detailed modeling of the clustering of halos hosting gas-rich galaxies, based on the $M_{HI}$-$M_h$ relation inferred from abundance matching. The results confirm our previous findings, by favoring an $M_{HI}$-$M_h$ relation that has large scatter and a weak dependence of $M_{HI}$ on host halo mass (see Figures \ref{fig:mh1mh} \& \ref{fig:2pcf_modelHI}). Lastly, we consider the consider the clustering of halos with different spin parameters. We find that halos with low spin parameters (as measured by the Bound-Density-Maxima halo finder algortihm) cluster more strongly than halos with higher spin parameters. Remarkably, this leads to the correlation function of ALFALFA galaxies being reproduced fairly well by just excluding low-spin halos from the computation (Figure \ref{fig:corr_spinbins}). This finding provides indirect support to the hypothesis that halo spin plays a central role in determining the gas contents of present-day galaxies.

\acknowledgements
\noindent
\textbf{Acknowledgements}

\noindent
The authors acknowledge the work of the entire ALFALFA collaboration team in observing, flagging, and extracting the catalog of galaxies used in this work. The ALFALFA team at Cornell is supported by NSF grants AST-0607007 and AST-1107390 to RG and MPH and by grants from the Brinson Foundation. EP would also like to thank Anatoly Klypin and Peter Behroozi for sharing their Bolshoi halo catalogs and for helpful comments. 

\noindent
Funding for the SDSS and SDSS-II has been provided by the Alfred P. Sloan Foundation, the participating institutions, the National Science Foundation, the US Department of Energy, the NASA, the Japanese Monbukagakusho, the Max Planck Society and the Higher Education Funding Council for England. The SDSS Web Site is \texttt{http://www.sdss.org}. The SDSS is managed by the Astrophysical Research Consortium for the participating institutions. 



\end{document}